\documentclass[11pt,a4paper,notitlepage]{article}
\usepackage[utf8]{inputenc}
\usepackage[english]{babel}
\usepackage[T1]{fontenc}
\usepackage{hyperref}
\usepackage{amsmath}
\usepackage{amsfonts}
\usepackage{color} 
\usepackage{caption}
\usepackage{array,multirow,makecell}
\setcellgapes{1pt}
\makegapedcells
\newcolumntype{R}[1]{>{\raggedleft\arraybackslash }b{#1}}
\newcolumntype{L}[1]{>{\raggedright\arraybackslash }b{#1}}
\newcolumntype{C}[1]{>{\centering\arraybackslash }b{#1}}
\newcommand{\Tr}{\mathrm{Tr}}

\newtheorem{definition}{Definition}
\newtheorem{proposition}{Proposition}
\newtheorem{corollary}{Corollary}

\newtheorem{lemma}{Lemma}
\usepackage{enumerate}
\usepackage{subfig}

\newcommand{\cG}{{\mathcal G}}

\newcommand{\cR}{{\mathcal R}}

\newcommand{\cT}{{\mathcal T}}

\newcommand{\cZ}{{\mathcal Z}}

\definecolor{mygray}{gray}{0.3}

\newcommand\beq{\begin{equation}}
\newcommand\eeq{\end{equation}}
\newcommand{\bes}{\begin{eqnarray}}
\newcommand{\ees}{\end{eqnarray}}
\def\nn{{\nonumber}}

\def\omegab{\overline{\omega}}
\def\psib{\overline{{\psi}}}

\newcommand\restr[2]{{
  \left.\kern-\nulldelimiterspace 
  #1 
  \vphantom{\big|} 
  \right|_{#2} 
  }}
\def\extd{\mathrm {d}}

\newcommand{\SU}{\mathrm{SU}}
\usepackage{multicol}
\usepackage{amssymb}
\usepackage{graphicx}
\usepackage[left=2cm,right=2cm,top=2.5cm,bottom=2.5cm]{geometry}
\begin{document}
\begin{center}
\textbf{\Large{
Renormalizable Group Field Theory beyond melonic diagrams: an example in rank four}}
\vspace{15pt}

{\large Sylvain Carrozza$^{a,}$\footnote{\url{scarrozza@perimeterinstitute.ca}}, Vincent Lahoche$^{b,}$\footnote{\url{vincent.lahoche@labri.fr}}} and Daniele Oriti$^{c,}$\footnote{\url{daniele.oriti@aei.mpg.de}}

\vspace{10pt}

$^{a}${\sl Perimeter Institute for Theoretical Physics\\
 31 Caroline St N, Waterloo, ON N2L 2Y5, Canada\\
}
\vspace{3pt}

$^{b}${\sl LaBRI, Univ. Bordeaux\\
351 cours de la Lib\'{e}ration, 33405 Talence, France, EU\\
}
\vspace{3pt}

$^{c}${\sl Max Planck Institute for Gravitational Physics\\
Am Mühlenberg 1, 14476 Potsdam, Germany, EU\\
}

\end{center}

\vspace{5pt}

\begin{abstract}
\noindent  We prove the renormalizability of a gauge-invariant, four-dimensional GFT model on $\SU(2)$, whose defining interactions correspond to necklace bubbles (found also in the context of new large-N expansions of tensor models), rather than melonic ones, which are not renormalizable in this case.   The respective scaling of different interactions in the vicinity of the Gaussian fixed point is determined  by the renormalization group itself. This is possible because the appropriate notion of {\it canonical dimension} of the GFT coupling constants takes into account the detailed combinatorial structure of the individual interaction terms. This is one more instance of the peculiarity (and greater mathematical richness) of GFTs with respect to ordinary local quantum field theories. We also explore the renormalization group flow of the model at the non-perturbative level, using functional renormalization group methods, and identify a non-trivial fixed point in various truncations.
This model is expected to have a similar structure of divergences as the GFT models of 4d quantum gravity, thus paving the way to more detailed investigations on them.
\end{abstract}

\setcounter{tocdepth}{1}
\tableofcontents
\pagebreak

\section{Introduction}
All modern quantum gravity approaches seem to suggest that the standard notions of continuum spacetime and geometry are merely effective and not fundamental. New types of degrees of freedom replace continuum fields, including the metric field of General Relativity as well as matter fields, with these new degrees of freedom being necessarily of a more abstract, non-spatiotemporal nature \cite{Oriti:2013jga}. Group field theories (GFTs) \cite{Baratin:2011aa,Oriti:2014uga, Oriti:2011jm} identify them as purely combinatorial and algebraic structures: graphs labeled by group-theoretic data. They are in fact peculiar quantum field theories on group manifolds (not identified as spacetime themselves) whose interactions are characterized by a non-local pairing of field arguments, in contrast with standard (relativistic) quantum field theories on spacetime where the requirement of locality of interactions translates into the conditions that fields appearing in the action are evaluated at the same point of the domain manifold. The type of the fundamental degrees of freedom that GFTs suggest is in common with canonical Loop Quantum Gravity \cite{Rovelli93, Oriti:2013aqa, Oriti:2014uga, Oriti:2014yla} as well as with the covariant spin foam models \cite{perez_review2012, bo,bo_bc, Reisenberger:2000zc}. The combinatorial structures of their interactions (which also dictates that of the graphs underlying the quantum states of the theory) is in common with tensor models \cite{Gurau:2011xq, Bonzom:2012hw, Gurau:2011xp, Gurau:2012vk, Gurau:2012vu, Gurau:2016cjo}, and even more so when the GFT interactions are chosen to satisfy a natural tensorial generalization of standard locality, namely \lq tensoriality\rq, to be defined precisely later in this paper. In this latter case, one often speaks of \lq Tensorial GFTs\rq. This results in GFT Feynman diagrams that are dual to simplicial (pseudo-)manifolds. Moreover, for many interesting GFT models, the associated Feynman amplitudes take the form of simplicial gravity path integrals, themselves the basis of independent approaches to quantum gravity \cite{bo_bc}. 

\

The nature of GFTs as quantum field theories brings with it a number of powerful technical tools and strategies for addressing open issues of these quantum gravity models, shared also by the related approaches. One is the need to constrain the inevitable ambiguities occurring in model building, i.e. in the very construction of candidate dynamical models of the fundamental building blocks of quantum spacetime. These include the starting group manifold, the rank of the basic (tensorial) GFT fields, the combinatorics to be used in their interaction terms, and the functional form of propagator and interaction kernels themselves. In addition, for models that are built starting from the quantization of some classical structure, one has to add the usual quantization ambiguities, i.e. the choice of the specific quantization map to be applied to it to obtain the quantum algebra of observables defining the theory. As in standard quantum field theories, one selection criterion is based on symmetry considerations. The study of symmetries in GFTs has been started recently \cite{BenGeloun:2011cz, Kegeles:2015oua, Kegeles:2016wfg}, but it has not yet gained a central role in constraining model building. The other main selection criterion is renormalizability, which becomes even more crucial when GFTs are intended to provide a fundamental (as opposed to effective only) definition of the quantum structure of spacetime, valid at all scales of energy and distance. Renormalization is the key tool to address the other outstanding issue of GFT models: (mathematical) control over their continuum limit, i.e. the regime of the quantum kinematics and dynamics in which many (potentially infinite) GFT degrees of freedom are simultaneously involved. It is in this regime that one expects a continuum (possibly smooth) spacetime manifold endowed with continuum fields to be a good approximation of the more fundamental GFT structures, and General Relativity to become a good approximation of their dynamics. Controlling the continuum limit of GFT models means controlling their non-perturbative renormalization group flow and mapping out their phase diagram. Both perturbative and non-perturbative renormalization are equally crucial research directions in other approaches to quantum gravity (see for example the work on spin foam models treated as (peculiar) lattice gauge theories \cite{Delcamp:2016dqo, Dittrich:2014mxa, Bahr:2012qj}). GFTs have the advantage that their renormalization can be tackled in a series of well-codified steps, adapted from those of standard quantum field theories, despite their more abstract nature of models for the pre-geometric building blocks of spaceetime.    

\

The main steps of GFT renormalization group studies are the following: 1) choose an initial theory space, possibly informed by symmetry considerations, and by results in related approaches, which suggest specific interpretations for the mathematical ingredients appearing in GFT models and their quantum amplitudes; 2) classify the divergences, setting up a precise power-counting; 3) truncate the model to power-counting renormalizable interactions; 4) understand how to reabsorb the perturbative divergences into some GFT analogue of "local" counter-terms (that is, more generally, in terms of interactions which are part of the theory space); when this is not possible, the initial theory space must be enlarged \footnote{This will not occur in the present paper, but we expect that it might in a proper 4d quantum gravity model. Whether an appropriate theory space can be found in this context is an open question which we shall keep for future investigations.}; 5) when this is possible, perturbative renormalizability at all orders follows, which can (for instance) be rigorously proven by means of multiscale methods; 6) the associated flow equations can be computed in the perturbative regime, elucidating the nature of the Gaussian fixed point. Depending on the model, asymptotic freedom may be realized; 7) one can then look for non-trivial fixed points, for example via functional renormalization group methods, through truncations of the Wetterich-Morris equation. These are the crucial elements to characterize the phase structure of the GFT model.  

\

These steps have been carried out, by now, for a large class of GFT models, and GFT renormalization is indeed a thriving area of developments, both at the perturbative level and, more recently, at the non-perturbative one, with most work dealing with tensorial GFTs (TGFTs). For a recent and almost up to date review, we refer to \cite{Carrozza:2016vsq}. A large class of GFT models has been shown to be perturbatively renormalizable, including both Abelian and non-Abelian ones \cite{BenGeloun:2011rc,Geloun:2012fq, Geloun:2013saa, Carrozza:2012uv, Carrozza:2013wda, Carrozza:2013mna, Samary:2012bw, Lahoche:2015ola}, with and without the additional (gauge-invariance) constraints on the GFT fields that characterize the corresponding Feynman amplitudes as lattice gauge theories and spin foam models. Asymptotic freedom has also been proven for many such renormalizable models, and seems to be a generic property in the simplest truncation of the theory space. Asymptotic safety has also been identified in a number of (more involved) TGFT models, in particular non-abelian ones with direct connection with 3d quantum gravity \cite{Carrozza:2014rya, Carrozza:2014rba, Carrozza:2016tih}. The non-perturbative GFT renormalization was also studied with many interesting results in the past few years. A few works adapted to GFTs the Polchinski equation \cite{Polchinski:1983gv, reiko_thomas1, reiko_thomas2, Krajewski:2016svb}, but most analyses adopted the functional renormalization group methods based on the Wetterich-Morris equation \cite{Benedetti:2014qsa, Geloun:2016qyb, Geloun:2016xep, Geloun:2016bhh, Lahoche:2016xiq, Benedetti:2015yaa}. This allowed to map the phase diagram of Abelian GFT models (in both compact and non-compact cases) and, more recently, non-Abelian ones, in appropriate truncations of the theory space. These analysis provided an independent confirmation of the results found with perturbative methods, and some important hints of possible phase transitions in GFT systems, in particular of condensation type, which in turn provide some indirect support for ongoing work on the extraction of cosmological dynamics from GFT models using condensate states \cite{Oriti:2016qtz, deCesare:2016rsf, Pithis:2016wzf, Gielen:2016dss, Gielen:2016uft, Oriti:2016ueo, Sindoni:2011ej}. All these analyses of the renormalization group flow of GFT models have relied heavily on the growing corpus of results obtained in the context of the simpler tensor models, which can be seen, in a sense, as GFTs stripped down of their group-theoretic data, leaving only the same combinatorial structures of both states and interactions, in particular those encoded in the Feynman diagrams of the theory. GFT renormalization has relied on results from tensor models first of all in the choice of the theory space (as mentioned, most analyses have dealt with TGFTs), but also in focusing the attention to truncations of the same theory space corresponding to so-called {\it melonic} interactions. This is motivated by the power counting of the analyzed TGFT models, but also by the fact that melonic diagrams have been shown to be the dominant ones in the large-N limit of tensor models [with the parameter $N$ of tensor models corresponding to the (large) mode cutoff of GFT models].

\

Such reliance on melonic interactions (thus melonic diagrams) raises one of the two main open issues in GFT renormalization: can it be overcome? can the GFT RG flow select instead other types of interaction processes for the GFT quanta? The issue is not merely technical, but has potentially important bearings on the physical interpretation of the same models. Doubts have been raised in fact on the physical significance of melonic interactions from a quantum gravity perspective: they correspond to spherical topologies, but to a very peculiar simplicial decomposition of the same, which seems to be  characterized by spectral and Hausdorff dimensions far from the topological one. This fact has been interpreted as a failure to reproduce a smooth spacetime in the continuum limit, and thus as a serious shortcoming for tensor models as models of quantum gravity, and has prompted a number of investigations aiming at exploring non-melonic regimes. Indirectly, this raises worries also for GFT models relying on the same combinatorial structures. The situation in GFT models of quantum gravity is more subtle than in tensor models, because any natural notion of distance would have to take into account the group-theoretic data as well as the combinatorial structure, while the definition of spectral and Hausdorff dimensions used in the context of tensor models were based on the simpler graph distance. Still, the issue of going beyond the melonic sector remains a very interesting one.  

The second main open issue in GFT renormalization, and possibly the most important one, is the analysis of GFT models of 4d quantum gravity with a more complete simplicial gravity interpretation. There are several of them, mainly based on the work done in the context of the spin foam formulation of loop quantum gravity (see for example \cite{bo_bc}). They are based on the Lorentz group manifold $\mathrm{SO}(3,1)$ (or on its Riemannian counterpart $\mathrm{SO}(4)$), and encode in their dynamics, in addition to the gauge-invariance conditions that have been successfully dealt with already in GFT renormalization analyses, further constraints ensuring the proper simplicial geometric interpretation of their quantum states and amplitudes. Such 4d quantum gravity models are obviously more technically involved and very little work has been done on them so far (but see \cite{Riello:2013bzw, Chen:2016aag, Bonzom:2013ofa}).
 
 \
 
In the present paper, we tackle the first issue above, and make indirectly some progress on the second one as well. More precisely, we study a simplified four-dimensional GFT model with $\SU(2)$ as base manifold, which:
on the one hand it is expected to have a similar structure of divergences as the 4d quantum gravity models, thus paving the way to more detailed investigations on them; on the other hand, it still remains in the class of gauge invariant models that we know how to renormalize in detail; that is, the generalized notion of locality provided by tensorial invariance is compatible with the (partial) geometric content of the model, as all the divergences can be expended in a basis of tensor invariant counter-terms. Most importantly, while remaining in the known class of gauge-invariant TGFT models, it goes beyond the melonic sector in a precise and very natural manner: contrary to all previously considered renormalizable TGFTs, in the present case the melonic interactions are {\it not} renormalizable; however, this does not prevent us from finding an appropriate class of interactions which do support a renormalizable theory: it is generated by so-called {\it necklace bubbles}, which have already been used to generate new $1/N$ expansions in the context of random tensor models \cite{Bonzom:2015axa, Bonzom:2016dwy}. In contrast to what happens in tensor models, however, no scaling is introduced by hand in our TGFT model, and it is (as it should) the renormalization group itself which defines the respective scaling of melonic and non-melonic bubbles in the vicinity of the Gaussian fixed point. At the technical level, the key insight that makes this possible is the realization that the appropriate notion of {\it canonical dimension} of the interaction coupling constants, for GFT models, has to take into account the detailed combinatorial structure of the individual interaction terms, and it is not simply dictated by the order of the same interactions. This is one more instance of the peculiarity (and greater mathematical richness) of GFTs with respect to ordinary local quantum field theories.

\

The paper is organized as follows. In Section 2, we introduce the 4d TGFT model to be renormalized. In Section 3, we prove the renormalizability to all orders in perturbation theory of the same model, restricted to necklace bubble interactions, up to order 6. In Section 4, we study the functional renormalization of the model, analyzing its deep UV regime (we will explain in the following what UV and IR mean in this GFT context), first, and the detailed properties of the Gaussian fixed point, in Section 5. We then move to a study of the non-perturbative regime of the model, thus away from the Gaussian fixed point, identifying and analyzing several non-trivial UV fixed points, first in the same order of truncation, in Section 6, and then going beyond this truncation towards higher orders, in Section 7, using additional combinatorial restrictions. In the final concluding section, we summarize our results.

\section{Four-dimensional tensorial GFT on SU(2)}\label{sec:model}

In this paper we are interested in the renormalizability and non-perturbative ultraviolet behavior of a TGFT in dimension four. A generic four-dimensional TGFT on $\SU(2)$ is defined by a partition function of the form
\begin{equation}\label{partitionfunction}
\cZ_\Lambda := \int \extd \mu_{C_\Lambda} [\psib,\psi] \, e^{-S_{\Lambda}[\bar{\psi},\psi]}\,,
\end{equation}
where $\psi (\textbf{g}) = \psi (g_1 , g_2 , g_3 , g_4 ) $ and $\bar{\psi} (\textbf{g})$ are complex fields over four copies of $\SU(2)$. The (UV regularized) Gaussian measure $d\mu_{C_\Lambda}$, which encodes the kinetic part of the classical action, is defined by the choice of the free propagator, which is:
\begin{equation}\label{propagator}
\int \extd \mu_{C_\Lambda}(\psi,\bar{\psi}) \, \psi(\textbf{g}) \bar{\psi}(\textbf{g}')= \int_{\SU(2)} \extd h \int_{1/\Lambda^2}^{\infty} \extd \alpha \, e^{-\alpha m^2(\Lambda)} \prod_{\ell=1}^4 K_{\alpha}(g_\ell h g_\ell^{\prime -1}),
\end{equation}
where $K_{\alpha}$ is the heat kernel on $\SU(2)$ at time $\alpha$, and $\Lambda > 0$ is an ultraviolet regulator\footnote{As is now standard in the GFT literature, short (resp. long) distances on the group are referred to as ultraviolet (resp. infrared). This nomenclature is essentially conventional, it has no bearing on the physical interpretation of the formalism. In particular, the correspondence between ordinary space-time scales and GFT scales cannot be established {\it a priori}, it will have to be inferred from the effective dynamics of the GFT fields, in the appropriate general relativistic limit.}. This UV regulator imposes a smooth cut-off on the large spin labels in the harmonic expansion of group functions, and then prevents any UV divergence. The global integration over the group manifold with the variable $h$ implements the \textit{closure constraint} (see \cite{Oriti:2013aqa, Krajewski:2012aw, Baratin:2011aa}), and provides the Feynman amplitudes with the structure of lattice gauge theories, discretized on a random lattice generated by the perturbative expansion itself. \\

The interaction part of the classical action $S_{\Lambda}$ is a weighted sum of (connected) tensorial trace invariants:
\beq
S_{\Lambda}[\psi, \psib]= \sum_{b} \lambda_b \, \Tr_b\left( \psi, \psib \right) 
\eeq
Each term in this sum is labeled by a \textit{bubble} $b$, which is a connected bipartite $4$-colored regular graph (more details on colored graphs and colored bubbles may be found in \cite{Ryan:2016emo} for instance). 
\begin{center}
\includegraphics[scale=.7]{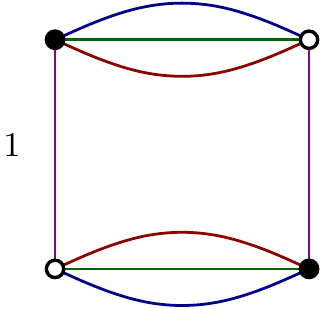} 
\captionof{figure}{Example of $4$-bubble}\label{figexample}
\end{center}
An example is pictured on Figure \ref{figexample}, and the correspondence between bubble and interaction is the following. Black and whites nodes correspond respectively to fields $\psi$ and $\psib$, whereas lines are delta functions identifying the corresponding group variable between two conjugates fields. Finally, all the group variables are integrated over the group manifold with respect to the Haar measure. Then, for the graph in Figure \ref{figexample}, the corresponding interaction writes:
\begin{equation}
 \Tr_{Fig \ref{figexample}} \left( \psi, \psib \right) := \int d\textbf{g} d\textbf{g}^\prime \psi(g_1,g_2,g_3,g_4)\bar{\psi}(g_1,g_2^\prime, g_3^\prime,g_4^\prime)\psi(g_1^\prime,g_2^\prime,g_3^\prime,g_4^\prime)\bar{\psi}(g_1^\prime,g_2, g_3,g_4)\,.
\end{equation}
At this stage, all the tensorial interactions may be allowed in the classical action. The renormalizability of the corresponding statistical model will be discussed in the next Section and will drastically reduce the set of relevant interactions. \\

The Schwinger $N$-point function, say $\mathsf{S}_N$, is formally defined by its perturbative expansion in power of the couplings $\lambda_b$, indexed by Feynman diagrams $\mathcal{G}$:
\begin{equation}
\mathsf{S}_N:= \sum_{\mathcal{G}\vert N(\mathcal{G})=N}\bigg(\prod_{b\in \mathcal{G}} (-\lambda_b)\bigg)\frac{1}{s(\mathcal{G})} \mathcal{A}_{\mathcal{G}}\,
\end{equation} 
where $N(\mathcal{G})$ denotes the number of external legs of the graph $\mathcal{G}$, $s(\mathcal{G})$ is a symmetry factor, and $\mathcal{A}_{\mathcal{G}}$ the Feynman amplitude of the diagram $\mathcal{G}$. The Feynman diagrams are then $5$-colored bipartite regular graphs, where the lines of colors $1, \cdots, 4$ are localized in the vertices, whereas the lines of color $0$ correspond to propagator lines. Figure \ref{Feynmanex} provides a typical example of a Feynman diagram, with three vertices and propagator lines pictured by dashed lines. Note that, in contrast to bubbles, Feynman diagrams may have open half-legs of color $0$. 
\begin{center}
\includegraphics[scale=1]{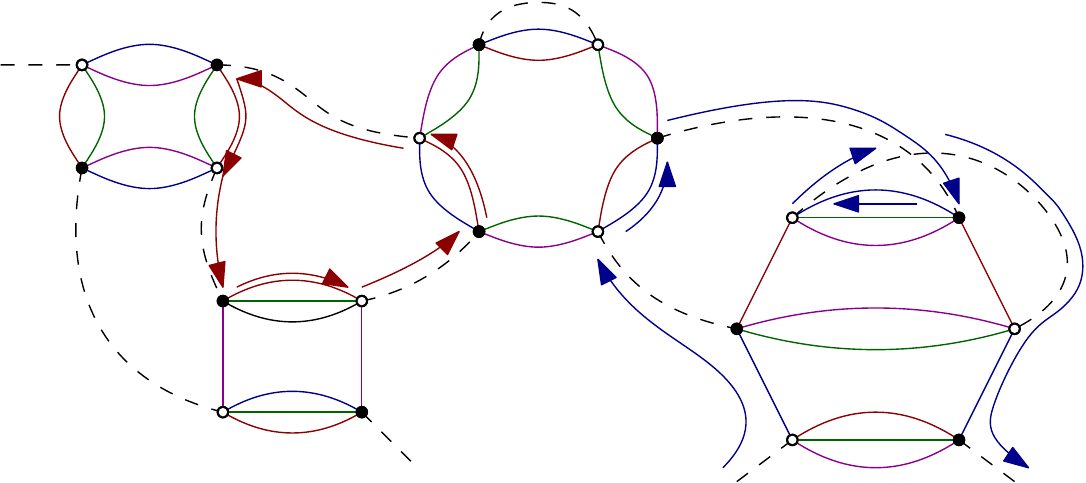} 
\captionof{figure}{A typical Feynman diagram occurring in the perturbative expansion. Blue arrows indicate an open face of color "0-blue" whereras red arrows indicate an internal face of color "0-red".}\label{Feynmanex}
\end{center}
Among the interesting features of Feynman diagrams in TGFTs, one can note that they are naturally equipped with a notion of \textit{face}, defined as a maximal and bicolored connected subset of lines, necessarily including the color $0$. A face may be \textit{closed} (or \textit{internal}) when the bicolored connected set corresponds to a cycle, and \textit{open} (or \textit{external}) otherwise (see Figure \ref{Feynmanex}). In this way, Feynman diagrams may be alternatively interpreted as $2$-complexes, that is gluings of faces, edges and nodes. Moreover, the colored structure of the diagram allows to unambiguously reconstruct a four-dimensional triangulation dual to a given colored $2$-complex. Then, a generic Feynman amplitude takes the form of a discretized lattice gauge theory on the associated $2$-complex, and can be expressed explicitly using standard properties of heat kernels as:
\begin{equation}\label{ampl_bulk}
\mathcal{A}_{\mathcal{G}}= \prod_{l\in L(\mathcal{G})} \int_{1/\Lambda^2}^{+\infty} d\alpha_l e^{-\alpha_l m^2}\prod_{f\in F(\mathcal{G})}K_{\alpha(f)} \bigg(\vec{\prod_{l\in f}} h_l\bigg)\prod_{f\in F_{ext}(\mathcal{G})}K_{\alpha(f)} \bigg( g_{s(f)}\vec{\prod_{l\in f}}h_lg_{t(f)}^{-1}\bigg)\,,
\end{equation}
where $L(\mathcal{G})$, $F(\mathcal{G})$ and $F_{ext}(\mathcal{G})$ are respectively the sets of propagator lines, internal and external faces in the diagram $\mathcal{G}$ and $s(f)$ (resp. $t(f)$) the index for source (resp. target) for the ends group variables of the external face $f$. Moreover, $\alpha(f):=\sum_{l\in f} \alpha_l$, $K_\alpha$ is the heat kernel on $SU(2)$ at time $\alpha$ and $\vec{\prod_{l\in f}} h_l$ is the oriented product of group elements around the face $f$. \\

For convenience for the rest of this paper, we introduce a projector onto the \textit{gauge invariant fields subspace}, $\hat{P}$, defined by its action on the fields:
\begin{equation}
\hat{P}\psi(g_1,g_2,g_3,g_4):= \int_{\SU(2)} dh \psi(g_1h,g_2h,g_3h,g_4h)\,,
\end{equation}
such that $\hat{P}\psi(g_1h,g_2h,g_3h,g_4h)=\hat{P}\psi(g_1,g_2,g_3,g_4)\,\forall h\in \SU(2)$. \\

The GFT we have just introduced cannot be expected to describe quantum gravity in four dimensions, but it is an interesting toy-model on our way towards this longer-term objective. Indeed, we anticipate its power-counting to closely mimic that of Euclidean quantum gravity models based on Barrett-Crane simplicity constraints \cite{bo_bc}, and possibly also that of models including the Immirzi parameter \cite{bo}. Such models are defined on the larger group $\mathrm{Spin}(4)= \SU(2) \times \SU(2)$, but the amplitudes have support on a quotient $\mathrm{Spin}(4) / \SU(2)$ by a diagonal $\SU(2)$ subgroup, or similarly defined subspaces of the Spin(4) group, depending on the Immirzi parameter. Henceforth, we expect that the exact scalings of the present  model will at the very least hold as bounds in the quantum gravity case. The key additional challenge we will have to face in quantum gravity models will be to understand how to reabsorb the divergences into appropriate counter-terms. As a consequence of the presence of simplicity constraints, Barrett-Crane spin foam amplitudes do not take the form of lattice gauge theories, they are genuine generalizations thereof. Hence, the clear relationship between the topology of the Feynman diagrams and the geometric support of their divergences\footnote{Namely, that the divergences of spherical amplitudes have support on trivial bulk holonomies up to gauge.} we will rely on in the present paper will need to be revised.  
\\

The power-counting of tensorial GFTs with gauge invariance condition has been largely studied in the recent past (see \cite{Carrozza:2012uv, Carrozza:2013wda, Carrozza:2013mna}) and is well understood. The so-called \emph{Abelian degree of divergence} of a Feynman diagram $\cG$ is defined as:
\beq
\omega (\cG ) := - 2 L(\cG) + 3 \left( F(\cG) - R(\cG) \right) \,,
\eeq
where $R(\cG)$ is the rank of the incidence matrix $\epsilon_{lf}$, whose entries are $+1$ (resp. $-1$) if $l$ belongs to $f$ and has the same (resp. opposite) orientation as $f$, and $0$ otherwise. For any graph $\cG$, $\omega(\cG)$ provides an upper bound on the \emph{superficial degree of divergence of $\cG$}, and the two degrees are equal when $\cG$ is spherical. This is in particular the case for \emph{melonic graphs}, which define a particular class of spherical triangulations. It is then easy to see that melonic interactions are not power-counting renormalizable in dimension four. Indeed, the recursive structure of the melonic diagrams ensures that \cite{Carrozza:2013wda}:
\beq
F(\cG)- R (\cG)= ( d - 2 ) \left(L(\cG)- V(\cG) + 1 \right) = 2 \left(L(\cG)- V(\cG) + 1 \right) \,,
\eeq 
where $d$ is the rank of the fields and $V(\cG) $ the number of interaction bubbles in $\cG$. From this result,  we deduce :
\beq
\omega(\cG) = 6 - 2 N(\cG) + \sum_{k \geq 1} \left( 4 k - 6 \right) n_{2 k}(\cG)\,,
\eeq
where $n_{2k}(\cG)$ denotes the number of vertices with $2k$ black and white nodes in $(\cG)$. Interestingly, this divergence degree is unbounded for any $N(\cG)$, as long as interactions with valency $4$ or higher ($k\geq 2$) are allowed. For instance, melonic graphs with only four-valent vertices verify:
\beq
\omega(\cG) = 6 - 2N(\cG) + 2 n_4(\cG)\,, \label{powercountingmelo}
\eeq
which cannot be uniformly bounded in $n_4(\cG)$. As a consequence, melonic divergences cannot be subtracted with a finite set of counter-terms, and it would therefore seem that the theory is non-renormalizable. However, this conclusion relies on the assumption that at least one melonic interaction is non-zero, and may not hold in full generality. In fact, we will see in the next section that this model is perturbative renormalizable if the theory space is restricted to a proper subset of tensorial bubbles, which is generated by so-called \textit{necklace bubbles}. The reason behind this somewhat surprising conclusion is that, unlike in ordinary local field theories, in GFT \emph{the scaling dimension of interactions is not a function of the valency alone}. Hence, the fact that melonic $\psi^4$ interactions are non-renormalizable is not sufficient to exclude the existence of other renormalizable bubbles. We will see explicitly that other types of four-valent interactions are indeed renormalizable. 

\section{Renormalizability in the necklace theory space}\label{sec:ren_neck}

The purpose of this section is to find a suitable theory space in which to renormalize our model. To this effect, we will first elaborate on the renormalizability of a given bubble interaction, as captured by the sign of its \emph{canonical dimension} (Definition \ref{def_cano_dim}). Crucially, we will point out that in GFT, such a quantity explicitly depends on the theory space in which the bubble is embedded. This will prompt us to introduce a coarser notion, the \emph{bubble dimension} (Definition \ref{def_bubble_dim}), which will provide us with an upper bound on the canonical dimension, independently of the choice of theory space. We will then prove general combinatorial properties of this estimate, which will allow to identify our candidate theory space, generated by \emph{necklace bubbles} (Definition \ref{def_necklaces}). We will finally go on and explicitly check the renormalizability of our model in the necklace theory space. Our proof will rely on several intermediate results about the combinatorial properties of necklace bubbles and Feynman diagrams. This line of arguments will culminate in Corollaries \ref{corollary2} and \ref{boundconv}, as well as Proposition \ref{proposition2}: the former imply power-counting renormalizability, while the latter guarantees that the perturbative divergences have the same functional form as the bare interactions; taken together, they establish the renormalizability of our model.

\

In standard QFT, perturbative renormalizability is closely related to the scaling dimension of the coupling constants. If all the coupling constants entering the bare action have positive or vanishing dimensions, the number of $N$-point functions which suffer from divergences is finite. As a result, these divergences can be cured by inclusion of a finite number of counter-terms, leading to a super-renormalizable or just-renormalizable theory\footnote{A super-renormalizable quantum field theory is one with finitely many divergent diagrams, while a just-renormalizable model generates infinitely-many divergent Feynman amplitudes.}. 
In contrast, when the dimension is negative, an infinite number of correlation functions diverge, which requires the inclusion of an infinite family of counter-terms. Physically, this means that an infinite number of coupling constants would need to be measured in order to make any prediction, henceforth predictivity is effectively lost. In this case, the theory is said to be perturbatively non-renormalizable. 

\medskip
\noindent
Strictly speaking, the GFT coupling constants are dimensionless quantities, but they can nonetheless be attributed a meaningful scaling dimension \cite{Carrozza:2013mna, reiko_thomas2, Benedetti:2015yaa}. It should capture the scaling dependence of the quantum corrections in the UV cut-off $\Lambda$, as predicted by the power-counting theorem. 
\begin{definition}\label{def_cano_dim}
For any bubble $b$, one defines the \emph{canonical dimension} $d_b$ as the unique integer such that: the leading-order radiative corrections contributing to the renormalization of $\lambda_b$ scale like $\Lambda^{d_b}$ in the limit $\Lambda\to + \infty$. 
\end{definition}
\noindent
As an example, let us consider the quantum correction to the four-point \emph{melonic} correlation function. For simplicity we assume that only $4$-valent vertices occur in the amplitudes, but it is easy to check that our result remains unchanged if we include other interaction bubbles. The perturbative expansion of the melonic $4$-point function takes the form (for vanishing external momenta):
\begin{equation}
S_{4,\mathrm{melo}} (\lambda_{4,{\mathrm{melo}}})=-\lambda_{4,{\mathrm{melo}}} +\mathcal{O}({\lambda_{4,\mathrm{melo}}}^2)\,,
\end{equation}
where the terms in $\mathcal{O}({\lambda_{4,\mathrm{melo}}}^2)$ typically behave as:
\begin{equation}
\mathcal{A}_{\mathcal{G}} \sim \lambda_{4,{\mathrm{melo}}}^{n_4(\mathcal{G})} \Lambda^{\omega(\mathcal{G})}\,.
\end{equation}
Denoting by $[X]$ the canonical dimension of the quantity $X$, these two equations imply that:
\begin{equation}
[\lambda_{4,\mathrm{melo}} ] \geq n_4(\mathcal{G}) [\lambda_{4,\mathrm{melo}}] + \omega(\mathcal{G})  \qquad \Rightarrow \qquad \left(n_4(\mathcal{G}) -1 \right) [\lambda_{4,\mathrm{melo}}] \leq 2 - 2 n_4(\mathcal{G})\,,
\end{equation}
where we have used the expression of the divergence degree of a $4$-valent melonic graph $\cG$. Clearly, this equation holds for any melonic graph if and only if $[\lambda_{4,\mathrm{melo}}] \leq -2$, which is negative, in agreement with our observation of the non-renormalizability of the melonic sector. Moreover, since melonic diagrams are the most divergent ones, we actually have $[\lambda_{4,\mathrm{melo}}] = -2$. In the same way, one can show that the canonical dimension of a coupling $\lambda_b$ associated to a melonic bubble of valency $N_b$ is:
\begin{equation}
[\lambda_b] = 6-2N_b < 0 \qquad \forall N_b\geq 4 \,. 
\end{equation}
Note however that this result relies on the implicit assumption that $4$-valent melonic bubbles have been included in the bare action. We remark also that non-melonic bubbles with the same valency will be associated to a different power-counting, and therefore a different scaling dimension. We conclude that, in contrast to ordinary scalar field theory, the value of the canonical dimension of a tensorial coupling constant explicitly depends on: 1) the theory space in which it is embedded; 2) the combinatorial structure of the bubble graph, which encodes the non-local nature of the interaction on the group manifold $\SU(2)^4$.

For our purpose, it is important to be able to estimate the canonical dimension of a coupling prior to fixing the theory space. More precisely, we are after a general upper bound on the canonical dimension, allowing to identify the coupling constants which are non-renormalizable in \emph{any theory space}. This is achieved by the following estimate.
\begin{definition}\label{def_bubble_dim}
The \emph{bubble dimension} $\tilde{d}_b$ of a bubble $b$ is defined as:
\beq\label{def_can_dim}
\tilde{d}_b = 2 - \underset{r \in \cR_2 (b)}{\mathrm{max}} \omega( r )\,,
\eeq
where $\cR_2 (b)$ denotes the set of single-vertex $2$-point graphs on $b$.
\end{definition}

In this definition \textit{the set of single-vertex $2$-point graphs on $b$} is the set of all $2$-points graphs which possess a single vertex of type $b$, and no other vertex. Note that, because $\cR_2 (b)=\emptyset$ if $b$ is the $2$-valent bubble, its bubble dimension is $2$, and therefore its canonical dimension should be bounded from above by $2$. The latter is actually a necessary requirement for any renormalizable theory with Laplace kinetic operators. Indeed, a $2$-point function diverging as $\Lambda^{k}$ would generate Laplace operator counter-terms of order $\Lambda^{k - 2}$, leading to inconsistencies whenever $k>2$.

We can now consider the quantum corrections to the $2$-point function involving a single bubble $b$, and deduce that:
\beq
d_b + \underset{r \in \cR_2 (b)}{\mathrm{max}} \omega( r ) \leq 2 \qquad \Rightarrow \qquad d_b \leq \tilde{d}_b 
\eeq
Hence, as announced, the bubble dimension of $b$ provides an upper bound on the canonical dimension $d_b$, without referring to the other bubbles in the theory space. \\

In order to investigate the perturbative renormalizability of the model, it is convenient to work out some basic properties of the bubble dimension under elementary graphs moves, especially under \emph{dipole moves}. Let us recall the standard definitions of dipoles, dipole contraction, and connected sum of two bubbles.
\begin{definition}
Let $\mathcal{G}$ be a Feynman graph. A $k$-dipole in this graph is a dashed line joining two nodes $n$ and $\bar{n}$ connected by exactly $k-1$ additional colored lines. 
\end{definition}

\begin{definition}
Let $l$ be a dashed line between two nodes $n$ and $\bar{n}$ in a graph $\mathcal{G}$. The contraction (or dipole contraction) operation consists in the following:\\

\noindent
$\bullet$ Deleting the two nodes linked by $l$, together with the colored lines joining $n$ and $\bar{n}$.\\

\noindent
$\bullet$ Reconnecting the external colored half-lines following their respective colors.\\

\noindent
The resulting graph is then called $\mathcal{G}/\{l\}$. Figure \ref{figcondip} below illustrates the contraction procedure.
\end{definition}\label{def1}
\begin{figure}[!h]
\centering
\includegraphics[scale=1]{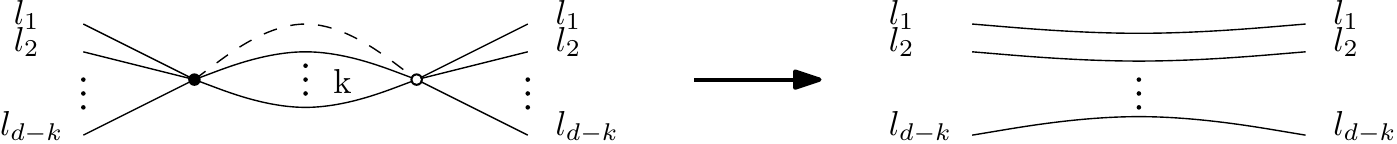} 
\caption{Contraction procedure of a line between two nodes joined with $k$ colored lines}\label{figcondip}
\end{figure}

\begin{figure}
\includegraphics[scale=1]{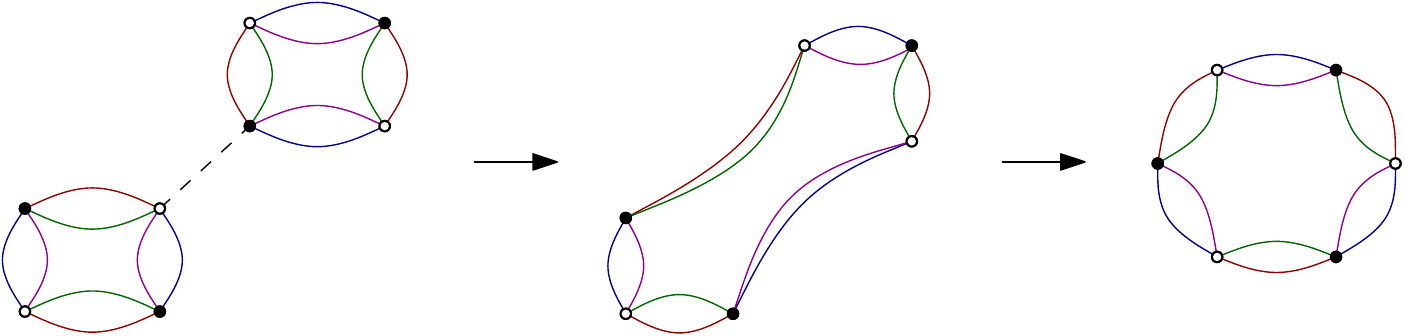} 
\caption{Connected sum of two elementary necklace bubbles.}
\label{sum}
\end{figure}

\begin{definition}
For any nodes $n_1 \in b_1$ and $\bar{n}_2 \in b_2$, the \emph{connected sum} $b_1 \sharp_{n_1, \bar{n}_2} b_2$ is the bubble obtained by first connecting $n_1$ and $\bar{n}_2$, and then contracting the newly created $1$-dipole. See Figure \ref{sum}. In particular, the sum of any bubble $b_1$ with the $2$-valent interaction $m_2$ gives the same bubble:
\begin{equation}
b_1 \sharp_{n_1, \bar{n}_2} m_2 = b_1 \quad \forall b_1, \forall n_1\in b_1,
\end{equation}
meaning that the $2$-valent interaction plays the role of the identity with respect to the sum $ \sharp$. 
\end{definition}\label{defsum}
We deduce the following lemma.
\begin{lemma}\label{con_sum}
Let $b_1$, $b_2$ and $b_3$ be three bubbles, such that $b_3$ is a connected sum of $b_1$ and $b_2$. Then:
\beq
\tilde{d}_{b_1} + \tilde{d}_{b_2} \geq \tilde{d}_{b_3} + 2\,.
\eeq
\end{lemma}
\textit{Proof}
$F$ and $R$ are both preserved by the contraction of a tree line\footnote{We call \emph{tree lines} any set of lines which do not form loops (and hence can be embedded into a tree). In particular, a tree line is a line which connects two distinct vertices.}, and are additive with respect to subgraphs which do not share internal faces. Therefore: 
\beq
\underset{r_1 \in \cR_2 (b_1)}{\mathrm{max}} (F-R)[r_1] + \underset{r_2 \in \cR_2 (b_2)}{\mathrm{max}} (F-R)[r_2] \leq \underset{r_3 \in \cR_2 (b_3)}{\mathrm{max}} (F-R)[r_3]\,,
\eeq
from which the lemma immediately follows. \hfill $\square$

\begin{figure}[!h]
\centering
\includegraphics[scale=0.5]{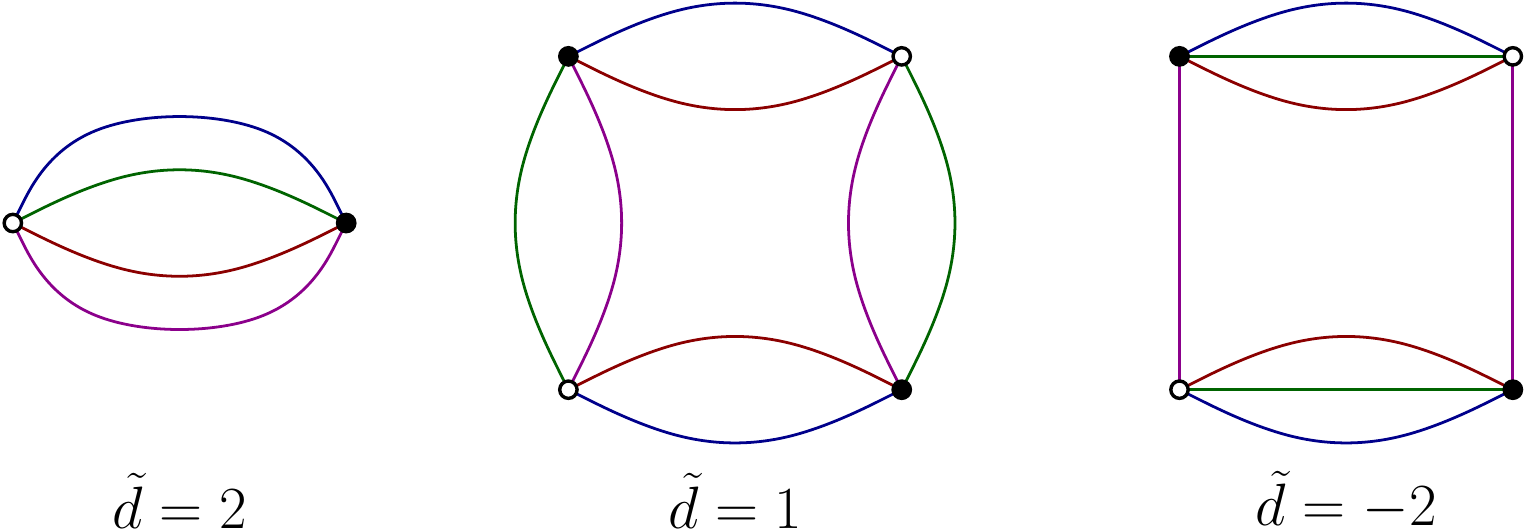} 
\caption{Connected invariants at order $2$ and $4$, and their bubble dimensions.}\label{phi2_4}
\end{figure}

\begin{figure}[!h]
\centering
\includegraphics[scale=0.5]{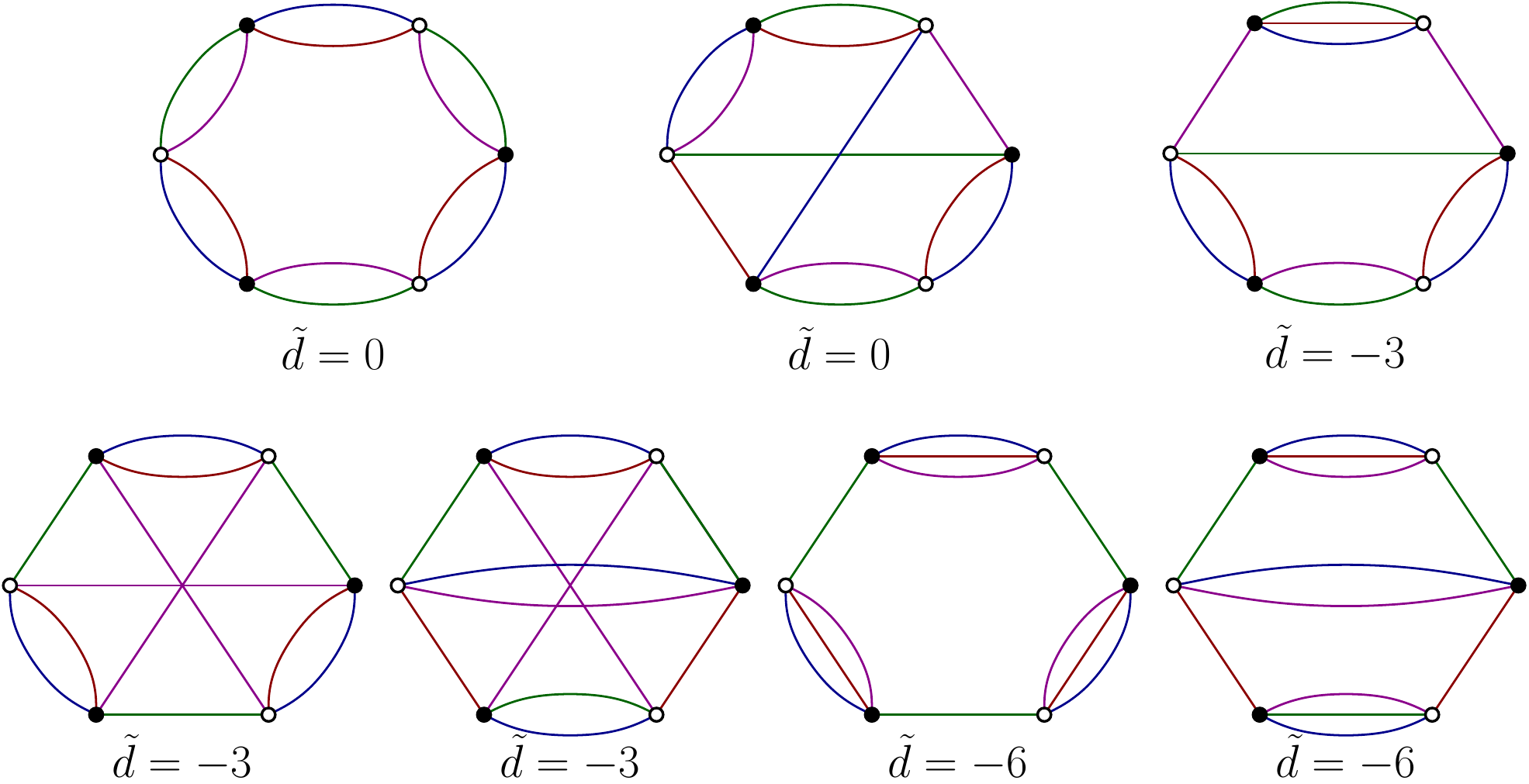} 
\caption{Connected invariants at order $6$ and their bubble dimensions.}\label{phi6}
\end{figure}

In order to define a renormalizable theory, it is instructive to investigate the bubble dimensions of the first few interactions. We expect bubbles with strictly negative canonical dimension to be irrelevant (or non-renormalizable). Let us start to list all possible connected operators. One has a unique term at order $2$, by definition of dimension $2$. At order $4$, there is one pattern (associated to three distinct colorings) with dimension $1$, which we call \emph{elementary necklace}. Note that each elementary necklace can be reconstructed from the knowledge of the color $i=2,3,4$ of the lines which are paired to the lines of color $1$. Accordingly, we will index each elementary necklace bubble with a double index $(1i)$. The other pattern is known as a \emph{melonic bubble} and has in this model a negative dimension $-2$. See Figure \ref{phi2_4}. At order $6$ we find seven patterns of contraction, shown in Figure \ref{phi6}, two of which are of dimension $0$ (top left corner), the five others being of strictly negative dimensions.

The two marginal $\psi^6$ invariants are easily obtained as connected sums of two elementary necklaces: two identical ones in the first case, and two different ones (associated to different splittings of the colors into two pairs of colors) in the second case, as illustrated in Figure \ref{phi6construction}. 
\begin{center}
\includegraphics[scale=0.7]{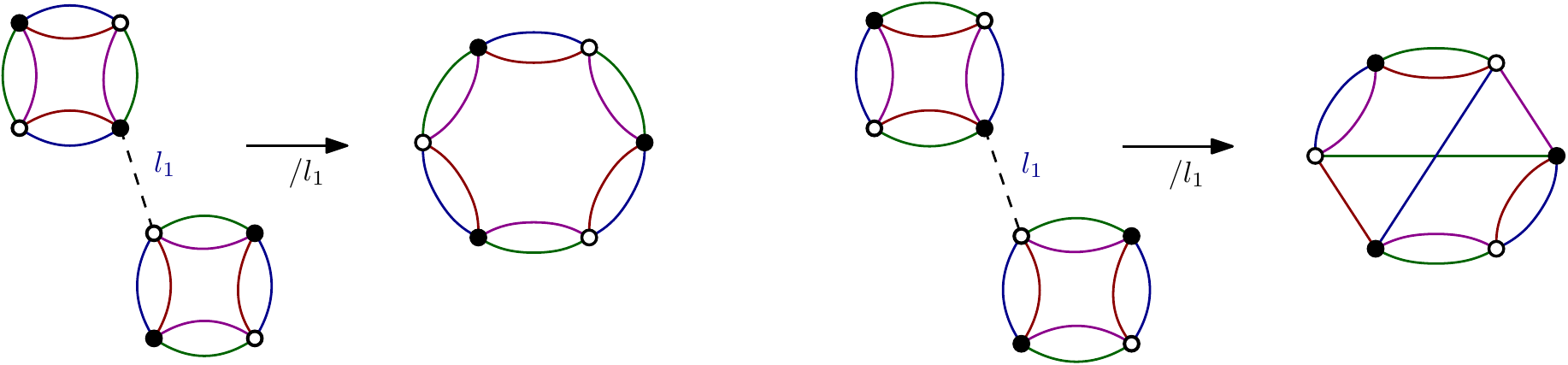} 
\captionof{figure}{Two typical contractions of elementary necklaces yield the two marginal $\psi^6$ invariants.}\label{phi6construction}
\end{center}
Moreover, by Lemma \ref{con_sum}, any other non-trivial connected sum of the invariants up to $\psi^6$ will result in strictly negative canonical dimensions. This indicates that, if consistent at all, a model having elementary necklaces as interactions will be renormalizable up to order $6$. In order to make this scenario more concrete, we gather a set of combinatorial results in the next subsection.

\medskip

\noindent{\bf{Remark.}} Note that we do not exclude the existence of positive dimension operators with higher valency, but we know that such operators cannot be generated at tree level by a theory with up to $\psi^6$ interactions. We will need to show in the following that this truncation remains consistent at any loop order. A similar question had to be addressed in the $\psi^6$ model of \cite{Carrozza:2013wda}, which features a $6$-valent interaction with strictly positive dimension\footnote{Namely, the bubble associated to this interaction is the complete bipartite graph $K_{3,3}$. In the sense of the present paper, the results of \cite{Carrozza:2013wda} immediately imply that it has a strictly positive canonical dimension, even though the notion of canonical dimension itself was not used in \cite{Carrozza:2013wda}.}. This extra coupling constant is however not generated by radiative corrections in the melonic sector, and can therefore be consistently set to $0$. 
\noindent

The previous computation provides intuition on the appropriate family of bubbles entailing a renormalizable theory. The family has to contain bubbles with positive or null canonical dimension, ensuring that the divergent degree can be uniformly bounded, and then that the divergences can be subtracted with a finite set of counter-terms. To be sure that the resulting theory space is stable, we moreover require the closure of the family under connected sums.

\begin{definition}\label{def_necklaces} The \emph{necklace bubbles} consist of the subset of bubbles generated by the $3$ elementary necklaces under the connected sum operation, and of the elementary $2$-valent bubble. We denote this subset by $\mathbb{B}$. 
\end{definition}
{\bf{Remark.}} The $2$-valent bubble is not essential, and only added for convenience.

Hence, any necklace bubble of valency higher than $4$ can be obtained by contraction of all the $1$-dipoles in a tree of elementary necklaces. It is then easy to see that any necklace bubble with valency higher than $4$ contains some $2$-dipole (each leave in the tree of elementary necklaces yields at least two such $2$-dipoles). The following lemma analyzes the effect of $2$-dipole contractions in a bubble.

\begin{figure}[!h]
\centering
\includegraphics[scale=0.9]{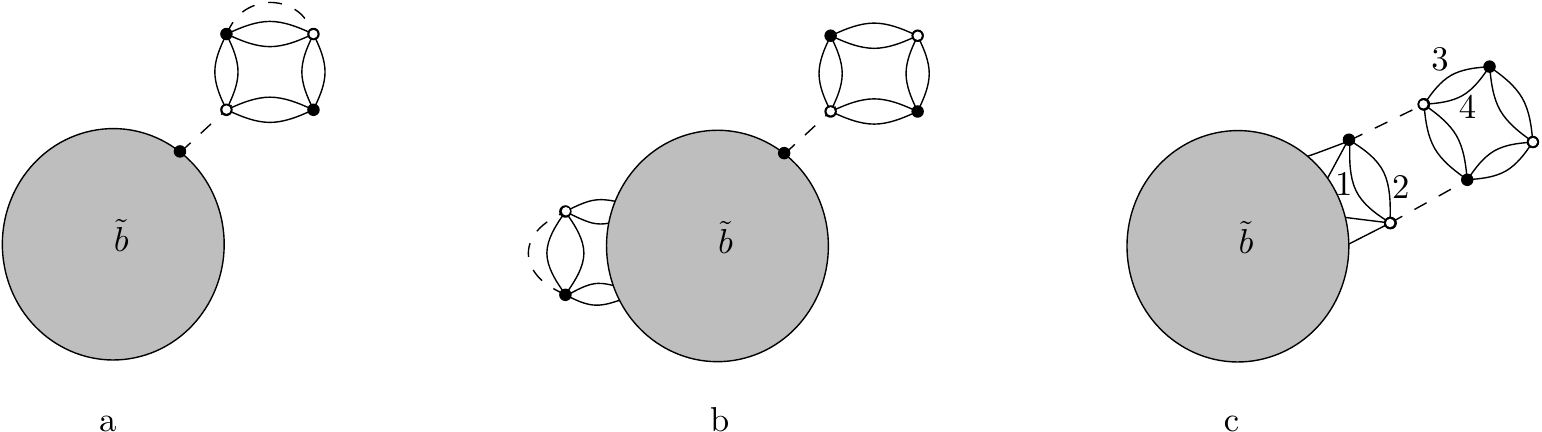} 
\caption{The three cases for the $3$-dipole contraction}\label{cases}
\end{figure}

\begin{lemma}\label{2_dipole_neck}
The set of necklace bubbles $\mathbb{B}$ is stable under contraction of $2$-dipoles. 
\end{lemma}
\textit{Proof}. 
Let $b$ be a necklace bubble. We can proceed by induction on $p = \frac{N_b}{2}$. The statement is empty when $p=1$. When $p=2$, $b$ is one of the three elementary necklaces, and the contraction of any $2$-dipole yields the $2$-valent bubble. Now assume $p \geq 3$. By definition, $b$ is the connected sum of an elementary necklace and a necklace bubble $\tilde{b}$ with $N_{\tilde{b}} = 2 (p - 1)$. We distinguish three cases pictured in Figure \ref{cases}. 
\begin{itemize}
\item \textit{Case \ref{cases}a} A $2$-dipole of the elementary necklace is contracted: the contraction of the $2$-dipole yields a $2$-valent bubble summed with the necklace $\tilde{b}$, which is nothing but $\tilde{b}$. \\

\item \textit{Case \ref{cases}b} A $2$-dipole of $\tilde{b}$ is contracted: by the induction hypothesis the $2$-dipole contraction yields a connected sum of a necklace bubble $\overline{b}$ and an elementary necklace, hence a necklace bubble.\\

\item \textit{Case \ref{cases}c} The contracted $2$-dipole corresponds to a line connecting $\tilde{b}$ and the elementary necklace. The only possibility is that $\tilde{b}$ and the elementary necklace share two faces of length $2$, and one easily checks the contraction of the two lines yields again bubble $\tilde{b}$, which is a necklace bubble.  
\end{itemize}
\hfill $\square$ 

\noindent{\bf{Remark.}} In particular, the contraction of $2$-dipoles in a necklace bubble conserves the connectedness of the bubble. 

\medskip

We now prove that the necklace theory space allows for renormalizability. Namely, we need to check that: 1) power-counting renormalizability holds; 2) the divergences are peaked around trivial bulk holonomies, ensuring a proper reabsorption of the divergences into 'local' counter-terms.

\begin{definition} A \emph{necklace graph} is a connected graph containing only necklace bubbles. The contraction of a spanning tree in a necklace graph with $L$ lines and $V$ bubbles yields a \emph{necklace rosette} with $L- V + 1$ loop lines.
\end{definition}
{\bf Remark.}
 By lemma \ref{2_dipole_neck}, the contraction of a $3$-dipole in a necklace graph yields another necklace graph, which is in particular connected. Moreover we recall that in full generality a rosette graph is a connected graph made out of a single vertex, which can in particular be obtained from the contraction of a spanning tree. 
 
\begin{definition}
For any necklace graph $\cG$, we define:
\beq
\rho(\cG) \equiv \left( L(\cG) - V(\cG) + 1 \right) - \left( F(\cG) - R(\cG) \right) \,.
\eeq
\end{definition}

\begin{lemma}\label{lemmainvariance}
$\rho$ is invariant under tree line contractions as well as $3$-dipole contractions. 
\end{lemma}
\textit{Proof}
$F$, $R$ and $L - V + 1$ (being the number of loops) are all invariant under contractions of tree lines.
Under a $3$-dipole contraction, $L$ and $F$ are changed into $L-1$ and $F-2$, while $R$ diminishes by exactly $1$. Hence $\rho$ is invariant under both moves. \hfill $\square$

\begin{definition}
 Let $\cG$ be a non-vacuum\footnote{Here and in what follows, we call \emph{non-vacuum} any Feynman graph with external legs.} necklace rosette with $L$ lines. We say that $\cG$ is \emph{$3$-dipole contractible} if there exists an ordering of its lines $l_1 , \ldots, l_L$ such that:
 \begin{itemize}
 \item $l_1$ is a $3$-dipole line of $\cG$; 
 \item For all $m \in \{ 2, \ldots , L \}$, $l_m$ is a $3$-dipole line in $\cG / \{ l_1, \ldots , l_{m-1}\}$.
 \end{itemize}
 \end{definition}
In words, a necklace rosette is contractible when all its lines can be successively contracted by means of $3$-dipole moves only.

\begin{figure}
\centering
\includegraphics[scale=0.5]{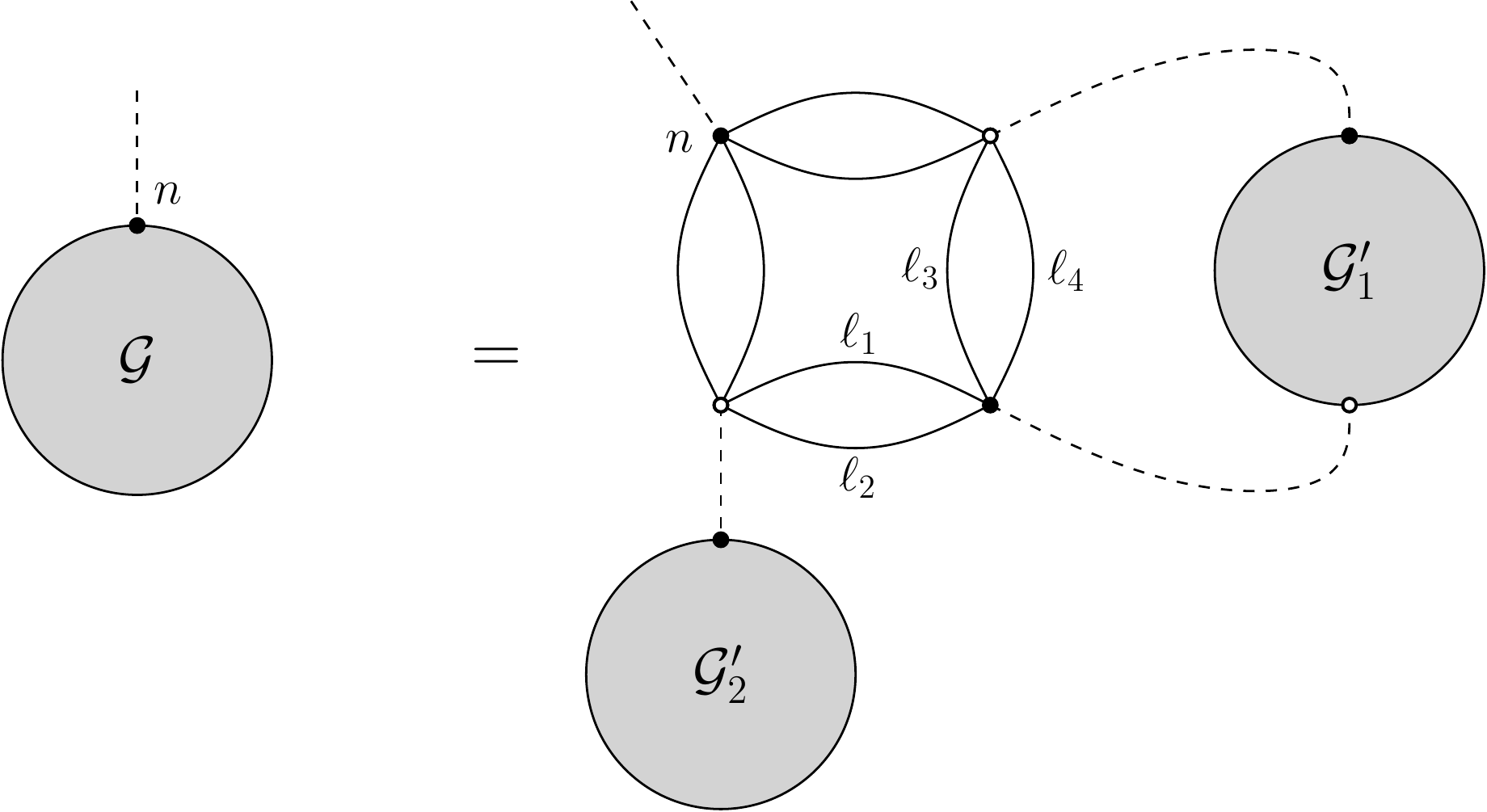} 
\caption{Structure of the bubble $b$ in the induction proving proposition \ref{bound_rho_4}.}\label{induction_lemma_rho}
\end{figure}
\begin{figure}
\centering
\includegraphics[scale=1]{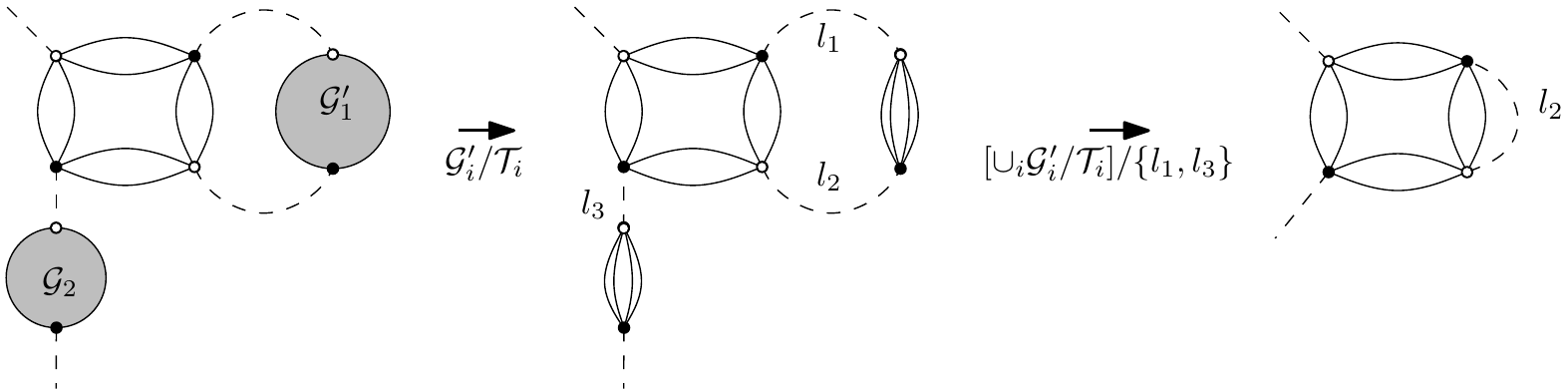} 
\caption{Illustration of the case $\Delta L=3$ and $\delta=2$.}\label{caseexample}
\end{figure}

\begin{proposition}\label{bound_rho_4} Let $\cG$ be a non-vacuum necklace graph with only $4$-valent vertices. Then:
\beq
\rho(\cG) \in \mathbb{N}\,,
\eeq 
and
\beq
\rho(\cG) = 0
\eeq
if and only if $\cG / \cT$ is $3$-dipole contractible for any spanning tree $\cT$ in $\cG$.
\end{proposition}
\textit{Proof.} Let us proceed by induction on the number of vertices $V$. When $V=1$, one easily checks that $\rho( \cG ) = 0$ and  the line of $\cG$ is a $3$-dipole. 
Assume now that $V\geq2$. Choose a vertex $v$ of $\cG$ such that $v$ is connected to an external leg of $\cG$ (say, at the node $n$). Call $\cG'$ the graph obtained from $\cG$ by deletion of $v$ and all the lines hooked to it. $\cG'$ has $\delta \leq 3$ connected components $\cG'_i$, $i = 1, \ldots, \delta$ (see Figure \ref{induction_lemma_rho} for $\delta = 2$). Each $\cG'_i$ is a non-vacuum necklace graph, therefore by the induction hypothesis:
\beq
\left( F- R \right) (\cG'_i) \leq \left( L - V \right) ( \cG'_i ) + 1 
\eeq 
and therefore 
\beq
\left( F- R \right) (\cG') \leq \left( L - V \right) ( \cG') + \delta\,. 
\eeq
We also have $V(\cG') = V(\cG) - 1$, hence denoting $\triangle L = L(\cG) - L(\cG')$ (one necessarily has $ \triangle L \geq \delta$):
\begin{equation}\label{ineq1}
\left( F- R \right) (\cG') \leq \left( L - V + 1 \right) ( \cG) + \delta - \triangle L \,. 
\end{equation}
Given the structure of the vertex $v$, which splits the faces between pairs of colors, one moment of reflection proves that:
\beq
\Delta( F - R) := \left( F- R \right) (\cG) - \left( F- R \right) (\cG') \leq \triangle L - \delta\,\label{ineq2}.
\eeq
This can be established by listing all cases, corresponding to between $1$ and $3$ external lines hooked to the vertex $v$. For instance, if $\Delta L = 3$ and $\delta=2$ as in Figure \ref{caseexample}, one must prove that $\Delta (F-R) \leq 1$. This is guaranteed by the fact that: if $\Delta F = 2$, then $\Delta R \geq 1$. Together, equations \eqref{ineq1} and \eqref{ineq2} yield $\rho( \cG ) \geq 0$.

\noindent
If $\cG$ admits a $3$-dipole contractible rosette, we can apply Lemma \ref{lemmainvariance} to show that $\rho( \cG ) = 0$. Reciprocally, assume that $\rho( \cG ) = 0$. Then one necessarily has:
\beq
\forall 1 \leq i \leq \delta , \, \left( F- R \right) (\cG'_i) = \left( L - V \right) ( \cG'_i ) + 1\,. 
\eeq
Hence for any $i$, by the induction hypothesis, for any spanning tree $\cT_i\in G'_i $, $\cG'_i / \cT_i$ is $3$-dipole contractible. Let $\cT$ be an arbitrary spanning tree in $\cG$. $\cT$ may be decomposed in a unique way as $\cup_{i=1}^\delta \cT_i \cup \{\ell_1,\cdots , \ell_{\delta}\}$, where $ \{\ell_1,\cdots , \ell_{\delta}\}\in \cG$ is a subset of lines hooked to $v$, and for any $i$, $\cT_i$ is a spanning tree in $\cG_i$. One can first contract the lines of the $\cT_i$ trees, and subsequently contract all the loops of $\cG'$ by $3$-dipole moves. The remaining $\triangle L$ lines form at most $\triangle L - \delta$ loops, each of which must bring $2$ faces in order to ensure $\left( F- R \right) (\cG) - \left( F- R \right) (\cG') = \triangle L - \delta$. There are three possible cases to examine: when there is no remaining loop (trivial); when there is $1$ remaining loop (then necessarily $\triangle L =3$ and $\delta = 1$, or $\triangle L = 3$ and $\delta = 2$); when there is $2$ remaining loops (then necessarily $\triangle L = 3$ and $\delta = 1$). It is easy to check that in each case one gets a $3$-dipole contractible graph after contraction of $\delta$ tree lines (see Figure \ref{caseexample} for the case $\Delta L=3$ and $\delta=2$). This achieves the proof. \hfill $\square$

\begin{corollary}\label{bound_rho_general} Let $\cG$ be a non-vacuum necklace graph. Then:
 \beq
 \rho(\cG) \in \mathbb{N}\,.
 \eeq
\end{corollary}
\textit{Proof.}
Replacing all the bubbles of $\cG$ by their corresponding trees of elementary necklaces does not change $\rho$, and one can therefore apply Proposition \ref{bound_rho_4}. \hfill $\square$ \\

\noindent
{\bf Remark.}
In particular, we can deduce that there cannot be any $3$-dipole in a necklace bubble, otherwise we could construct a necklace graph with a single $4$-dipole line, which would yield $\rho = -1$. This gives a feeling of why the necklace sector $\mathbb{B}$ is free from badly divergent melonic contributions.  

\begin{corollary}\label{corollary2} Any necklace bubble $b$ with valency $N_b$ has canonical dimension
\beq
d_b = 3 - \frac{N_b}{2}\,.
\eeq
\end{corollary}
\textit{Proof.}
By definition of the canonical dimension, this statement is equivalent to:
\beq
\underset{r \in \cR_2 (b)}{\mathrm{min}} \rho( r ) = 0\,.
\eeq
The inequality $\underset{r \in \cR_2 (b)}{\mathrm{min}} \rho( r ) \geq 0$ is guaranteed by Corollary \ref{bound_rho_general}. To prove that it is saturated, it suffices to note that necklace bubble always contain a $2$-dipole, and by Lemma \ref{2_dipole_neck} the contraction of such a $2$-dipole yields again a necklace bubble. Hence one can construct a $3$-dipole contractible $2$-point rosette on $b$, whose $\rho$ clearly vanishes. \hfill $\square$

For later computations of beta function, it is convenient to introduce a \emph{dimensionless degree of divergence}, by subtracting the scaling contributions of the bubbles from $\omega$. In the necklace sector of our model, the dimensionless power counting degree $\omegab$ is defined as:
\beq
\omegab := \omega - \sum_{b\in\mathbb{B}} n_b \, d_b 
\,\,.
\eeq
Elementary algebraic manipulations then allow to prove that:
\beq
\omegab = 3 - \frac{N}{2} - 3 \rho\,.
\eeq 

\noindent
As a consequence, relative to the correctly normalized degree of divergence $\omegab$, only the graphs with $\rho=0$ can be superficially divergent. Then, we call \textit{Leading Order} (LO) the non-vacuum Feynman graphs with $\rho=0$, and \textit{Non-Leading Order} (non-LO) the non-vacuum graphs for which $\rho>0$. As a direct consequence of the previous counting:
\begin{corollary}\label{boundconv}
In the necklace sector, any non-LO non-vacuum Feynman graph $\mathcal{G}$ verifies:
$$\omegab(\mathcal{G})\leq- N(\mathcal{G})/2.$$ 
\end{corollary}
Then, all the non-LO graphs are superficially convergent, and therefore, do not require renormalization, except for sub-
divergent diagrams. Renormalization requires a locality principle, that is a way to define contraction of high subgraphs into local counter-terms. In the TGFT context, the appropriate notion of locality is called \emph{traciality}, with the following definition \cite{Carrozza:2012uv, Carrozza:2013wda, Carrozza:2016tih}:
\begin{definition}
A connected Feynman graph $\mathcal{G}$ is said to be \emph{tracial} if the flatness of the bulk generalized connection (see equation \eqref{ampl_bulk}) imposes that it is trivial up to gauge.  
\end{definition}
It has been pointed out that melonic diagrams are tracial, and then, that renormalization makes sense for models for which all the divergent diagrams are melonics. The following proposition ensures that necklace graphs are also tracial, and thereby that the necklace sector is not only power-counting renormalizable but renormalizable, in the sense that we can define counter-terms and a systematic subtraction procedure of divergent subgraphs, which can be constructed rigorously using multiscale methods as for the melonic sector (see\cite{BenGeloun:2011rc,Geloun:2012fq, Geloun:2013saa, Carrozza:2012uv, Carrozza:2013wda, Carrozza:2013mna, Samary:2012bw, Lahoche:2015ola}).
\begin{proposition}\label{proposition2}
In the necklace sector, LO graphs are tracial.  
\end{proposition}
\textit{Proof.}
Let $\cG$ be a LO graph. One can choose an arbitrary spanning tree $\cT$ in $\cG$ and gauge fix the holonomies to the identity along $\cT$. Flatness imposes that the holonomy associated to any $3$-dipole in $\cG / \cT$ vanishes, and since $\cG  / \cT$ is $3$-dipole contractible, one may proceed by induction to show that all bulk holonomies are actually trivial. \hfill $\square$ \\

This last proposition guarantees that the necklace sector is not only power-counting renormalizable, but renormalizable. A rigorous proof at all orders in perturbation can for instance be constructed by means of multiscale methods (see e.g. \cite{Carrozza:2013wda}). 

\section{Functional renormalization group}\label{sec:frg}

\subsection{Formalism and approximation scheme}

In this section we move on to non-perturbative aspects of the renormalization group flow using the functional renormalization group (FRG) formalism, which has been successfully applied to TGFTs for several models. The reader may consult the references in the introduction for extended discussions, and \cite{Wetterich:1993ne, Morris:1993qb, Canet:2003qd, Berges:2000ew, Bagnuls:2000ae, Delamotte:2007pf, Rosten:2010vm, Gies:2006wv} for a general introduction to the FRG formalism in QFTs. 

The FRG is based on a one-parameter deformation of the original free energy :
\begin{equation}
e^{W_{\Lambda,k}[J,\bar{J}]}:=\int d\mu_{C_\lambda}(\bar{\psi}, \psi)e^{-S_\Lambda[\bar{\psi}, \psi]-\Delta S_k[\bar{\psi}, \psi]+(\bar{J},\psi)+(\bar{\psi},J)}
\end{equation}
where $(\bar{J},\phi):=\int d\textbf{g} \bar{J}(\textbf{g})\psi(\mathbf{g})$, $k\in[0,\Lambda]$ is the deformation parameter and $W_{\Lambda,k}[J,\bar{J}]$ the standard free energy. The deformation is induced by the new quadratic term $\Delta S_k[\bar{\psi}, \psi]$, depending on $k$ only
\begin{equation}
\Delta S_k[\bar{\psi}, \psi]:=\int d\textbf{g}_1d\textbf{g}_2 \bar{\psi}(\textbf{g}_1)R_k(\textbf{g}_1\textbf{g}_2^{-1})\psi(\textbf{g}_2)\,.
\end{equation}
The regulator $R_k(\textbf{g})$ only depends on the geodesic lengths $\vert g_i \vert$ for each component $g_i$ of $\textbf{g}$. We do not list the standard properties of the regulator here. We only mention that the boundary conditions of the function are chosen such that for $k=\Lambda$ the classical action $S_\Lambda$ becomes an UV boundary condition for the renormalization group flow (i.e. the regulator becomes very large for all modes), whereas the original generating functional is recovered in the complementary limit $k=0$, when all the fluctuations are integrated out (i.e. the regulator goes to zero for every modes). Between these boundaries, the shape of the function is chosen such that the UV modes are almost unaffected by the regulator term, whereas the IR modes acquire a large mass which decouples them from the UV dynamics. \\

\noindent
The dependence on $k$ may be captured by a first order differential equation, the so-called \textit{Wetterich-Morris} equation. This equation involves the (modified) Legendre transform $\Gamma_{\Lambda,k}$ of the free energy with respect to the source, that is to say, the \textit{effective average action}:
\begin{equation}
\Gamma_{\Lambda,k}[\bar{\phi},\phi]+\Delta S_k[\bar{\phi}, \phi]:= (\bar{J},\phi)+(\bar{\phi},J)-W_{\Lambda,k}[J,\bar{J}]\,,
\end{equation}
where the mean field $\phi$ is defined as the first functional derivative with respect to the conjugate external source $\phi=\delta W_{\Lambda,k}/\delta \bar{J}$. Deriving the effective action with respect to $k$, and taking into account that the mean field $\phi$ is a gauge invariant field ($\hat{P}\phi=\phi$), one arrives at the renormalization group flow equation:
\begin{equation}
\partial_k\Gamma_{\Lambda,k}[\bar{\phi},\phi]=\int d\textbf{g}_1d\textbf{g}_2d\textbf{g}_3 \partial_kR_k(\textbf{g}_1\textbf{g}_2^{-1})(\Gamma^{(2)}_{\Lambda,k}+R_k)^{-1}(\textbf{g}_2,\textbf{g}_3)\hat{P}(\textbf{g}_3,\textbf{g}_1)\,,\label{WM}
\end{equation}
where :
\begin{equation}
\Gamma^{(2)}_{\Lambda,k}:=\frac{\delta^2}{\delta \psi \delta\bar{\phi}}\Gamma_{\Lambda,k}\,.
\end{equation}
Physically, the renormalization group equation (RGE) \eqref{WM} describes the evolution of the effective action when high energy modes with respect to the running scale $k$ are integrated out from higher to lower scales; and interpolates between the classical action for $k=\Lambda$ and the full effective action $\Gamma_\Lambda$\footnote{This is nothing but the usual Legendre transform of the free energy, and the generating functional of the one-particle irreducible Feynman graphs.} in the complementary limit $k=0$. \\

\noindent
The RGE \eqref{WM} is exact, but as usual in physics with exact equations, extracting information requires an appropriate approximation scheme. There are several approximation procedures and refinements for each of them, largely discussed in the literature. For our purpose, we will only consider the most commonly used approximation, called \textit{truncation procedure}, and consisting in a systematic projection of the renormalization group flow into a finite-dimensional subspace of couplings. Any such finite-dimensional theory subspace translates into a particular ansatz for $\Gamma_{\Lambda,k}$. For the moment, let us consider the simplest truncation to perturbatively renormalizable interactions. In graphical notation for the interactions, this yields:
\begin{align}\label{ansatz_eff_action}
\nonumber\Gamma_k &= -Z(k) \sum_{\ell=1}^4 \vcenter{\hbox{\includegraphics[scale=0.8]{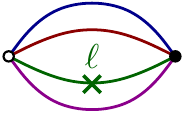}}}  + Z(k) k^2 u_2(k) \vcenter{\hbox{\includegraphics[scale=0.8]{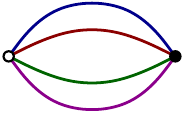}}}  +  Z(k)^2 k \frac{u_{4} (k)}{2}  \sum_{\ell=2}^4 \vcenter{\hbox{\includegraphics[scale=0.8]{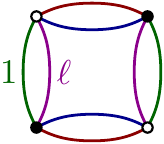}}}  \\
&\qquad + Z(k)^3 \frac{u_{6,1} (k)}{3} \sum_{\ell = 2}^4 \vcenter{\hbox{\includegraphics[scale=0.8]{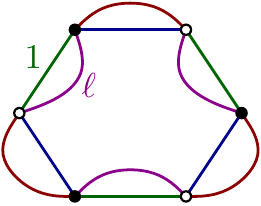}}} + Z(k)^3  u_{6,2} (k) \sum_{\substack{\ell_1 , \ell_2 =2\\ \ell_1\neq \ell_2}}^4 \vcenter{\hbox{\includegraphics[scale=0.8]{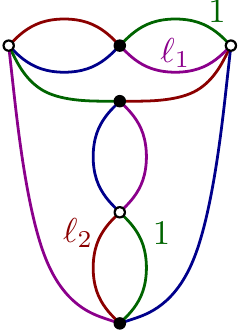}}}\,,
\end{align}   
where the cross on the green edge in the first term represents a Laplacian insertion $\Delta_\ell$. Note that this approximation scheme corresponds to the standard \textit{local potential approximation} in the FRG literature, tensoriality playing the same role as space-time locality in ordinary quantum field theory, and derivative couplings being discarded (except in the kinetic term). The ansatz \eqref{ansatz_eff_action} only involves five parameters $m^2$, $Z$ and $\{\lambda_b\}$, and depends on the running scale $k$. Moreover, since the effective average action reduces to the bare action at $k=\Lambda$, the boundary conditions for the coupling constants are consistently specified by the perturbative definition of the theory, and in particular $Z(k=\Lambda)=1$. \\

The approximation scheme depends on a choice of regulator function $R_k$. As long as we do not truncate the flow, the RGE equation \eqref{WM} is exact\footnote{It is “exact" as a formal functional equation.}. However, the truncation generally introduces a spurious dependence on the choice of the regulating function, and the reliability of the result obtained in a given truncation for a specific choice of the regulator is an important challenge of the FRG method, which is not specific to the TGFT context (the reader may consult \cite{Litim:2001up, Codello:2013bra, Morris:1993qb} for a discussion on the “optimization problem"). In this paper, we will only consider the regulator chosen in \cite{Carrozza:2016tih}, in the context of a three-dimension TGFT on $SU(2)$, whose Peter-Weyl components are:
\begin{equation}
\hat{R}_k(\{j_l,m_l,n_l\}):=Z(k)k^2 r_{\Lambda,k}\bigg(\frac{\sum_{l=1}^4 j_l(j_l+1)}{k^2}\bigg)\,,
\end{equation}
where $j_l$ are the labels of the irreducible representations of $SU(2)$, $m_l$ and $n_l$ the magnetic indices, and the function $r_{\Lambda,k}$ is defined as:
\begin{equation}
r_{\Lambda,k}(z)=\frac{z}{e^{-zk^2/\Lambda^2}-e^{-z}}-z  \underset{k\ll \Lambda}{\sim}\frac{z}{e^z-1}\,.
\end{equation}
The choice of this function is essentially motivated by convenience for calculations. In particular:
\begin{equation}
\mathcal{C}_k^{-1}(\textbf{g}_1,\textbf{g}_2)=\bigg[-Z(k)\sum_{l=1}^4\Delta_l+R_k\bigg]^{-1}(\textbf{g}_1,\textbf{g}_2)=\frac{1}{Z(k)}\int_{\Lambda^{-2}}^{k^{-2}} d\alpha \prod_{l=1}^4K_\alpha(g_{1l}g_{2l}^{-1})\,.
\end{equation}

\subsection{Flow equation in the deep ultraviolet regime}

The approximation scheme being completely fixed, we are in position to extract the truncated flow equations for $m^2$, $Z$ and $\{\lambda_b\}$ from the full RGE equation \eqref{WM}. The derivation is based on an expansion of the right-hand side of the RGE equation \eqref{WM} in powers of fields, and an identification term by term with the left-hand side. The contributions on the right-hand side may be represented in a diagrammatic way as one-loop Feynman graphs with two types of propagators. One of  them is
\begin{equation}
\mathcal{K}_k^{-1}(\textbf{g}_1,\textbf{g}_2):=[ \mathcal{C}_k+Z(k)m^2(k)]^{-1}(\textbf{g}_1,\textbf{g}_2)\,,
\end{equation}
which may occur many times in a single diagram. The second propagator can be written (in a matrix notation for the product) as
\begin{equation}
D_k(\textbf{g}_1,\textbf{g}_2):=\hat{P}\mathcal{K}_k^{-1}\partial_kR_k\mathcal{K}_k^{-1}\hat{P}(\textbf{g}_1,\textbf{g}_2)\,,
\end{equation}
and appears exactly once in any given diagram. Note that both $\mathcal{K}_k^{-1}$ and $D_K(\textbf{g}_1,\textbf{g}_2)$ may be expressed as a product of heat kernels $K_\alpha$, one for each component, and then only depends on $\textbf{g}_1\textbf{g}_2^{-1}$. In the rest of this paper, we will use extensively a diagrammatic representation for the flow equations, with the convention that dotted lines represent contractions with the propagator $\mathcal{K}_k^{-1}$ and dashed lines represent contractions with the second propagator $D_k$. An example of such diagram, representing a contribution involving six fields, is pictured on Figure \ref{examplediagram} below. 
\begin{center}
\includegraphics[scale=1]{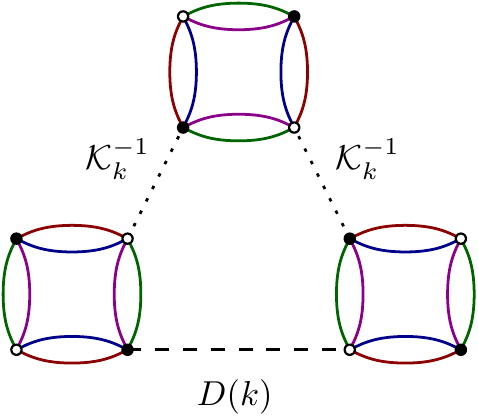} 
\captionof{figure}{A typical contribution of the right-hand side of the Wetterich-Morris equation involving six fields, and a single loop of length three.}\label{examplediagram}
\end{center}
In this paper, we will only consider the ultraviolet regime of the theory, corresponding to:
\begin{equation}
1\ll k\ll \Lambda\,.
\end{equation}
Note that imposing $k\ll \Lambda$ is equivalent to imposing the limit $\Lambda \to \infty$, in which the RGE equation \eqref{WM} is well defined, in contrast to the original partition function. Then, we will assume that this limit has been taken once and for all, and will remove the index $\Lambda$ from our equations. The restriction to the ultraviolet regime ensures many simplifications, and in particular, only some contractions in the right-hand side of the Wetterich--Morris equation provide a relevant contribution to the renormalization group flow. Indeed, we will see that all these contributions, maximizing the number of internal faces, come from the one-loop \textit{necklace diagrams}, which generate effective necklace bubbles. This is guaranteed by the traciality of necklace diagrams, and in turn implies the stability of the necklace theory space under the renormalization group flow.\\

\noindent
Equations \eqref{flow_graph_u2} to \eqref{flow_graph_u62} provide, in a pictorial way, the leading contributions in the ultraviolet regime to the relevant parameters in our truncation.
\beq\label{flow_graph_u2}
\partial_k ( Z(k) k^2 u_2(k) ) \, \vcenter{\hbox{\includegraphics[scale=0.8]{Figures/int2.pdf}}} - \partial_k Z(k) \sum_{\ell=1}^4 \, \vcenter{\hbox{\includegraphics[scale=0.8]{Figures/int2_laplace.pdf}}} \approx - Z(k)^2 k \, u_{4} (k)  \sum_{\ell=2}^4 \left(  \vcenter{\hbox{\includegraphics[scale=0.8]{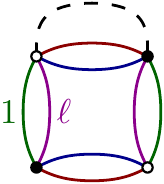}}} +  \vcenter{\hbox{\includegraphics[scale=0.8]{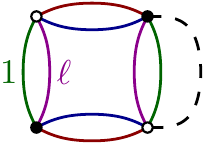}}} \right)\,,
\eeq

\begin{align}\label{flow_graph_u4}
\partial_k \left( Z(k)^2 k \frac{u_4 (k)}{2} \right) & \sum_{\ell=2}^4 \vcenter{\hbox{\includegraphics[scale=0.8]{Figures/int4_l.pdf}}} \approx - Z(k)^3 u_{6,1} (k) \sum_{\ell = 2}^4 \left( \vcenter{\hbox{\includegraphics[scale=0.8]{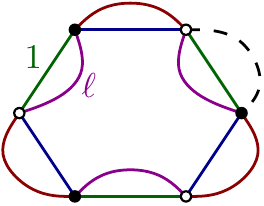}}} + \vcenter{\hbox{\includegraphics[scale=0.8]{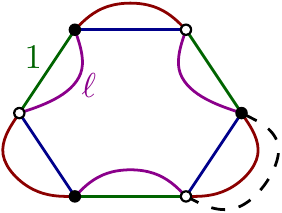}}} \right) \nn \\
& - Z(k)^3 u_{6,2} (k)  \sum_{\substack{\ell_1 , \ell_2 =2\\ \ell_1\neq \ell_2}}^4  \left( \vcenter{\hbox{\includegraphics[scale=0.8]{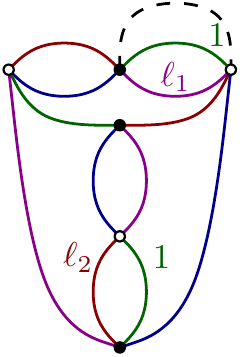}}} + \vcenter{\hbox{\includegraphics[scale=0.8]{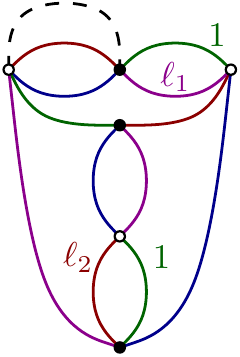}}} +  \vcenter{\hbox{\includegraphics[scale=0.8]{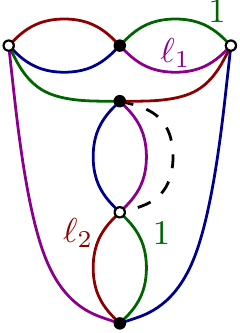}}} +  \vcenter{\hbox{\includegraphics[scale=0.8]{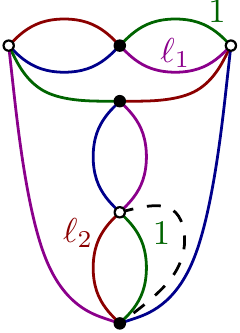}}} \right)\\\nonumber
&+ \left( Z(k)^2 k u_4 (k) \right)^2 \, \sum_{\ell=2}^4 \left(  \vcenter{\hbox{\includegraphics[scale=0.8]{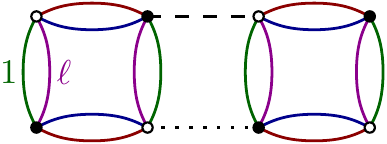}}} +   \vcenter{\hbox{\includegraphics[scale=0.8]{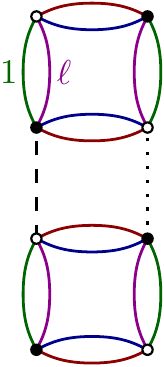}}} \right) \nn\,,
\end{align}

\begin{align}\label{flow_graph_u61}
\partial_k \left( Z(k)^3 \frac{u_{6,1} (k)}{3} \right) & \sum_{\ell = 2}^4 \vcenter{\hbox{\includegraphics[scale=0.8]{Figures/int61_l_small.pdf}}} \approx 2 Z(k)^5 k u_4(k) u_{6,1}(k) \sum_{\ell = 2}^4 \left( \vcenter{\hbox{\includegraphics[scale=0.8]{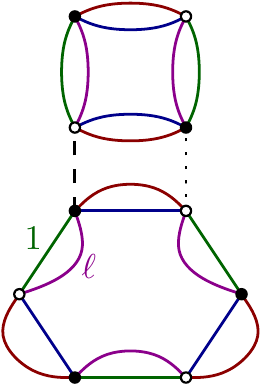}}} + \vcenter{\hbox{\includegraphics[scale=0.8]{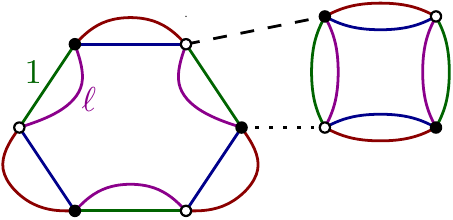}}} \right) \\\nonumber
&-  \left( Z(k)^2 k u_4(k) \right)^3 \sum_{\ell = 2}^4 \left( \vcenter{\hbox{\includegraphics[scale=0.8]{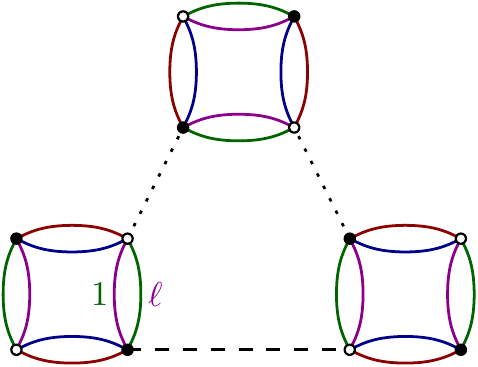}}} + \vcenter{\hbox{\includegraphics[scale=0.8]{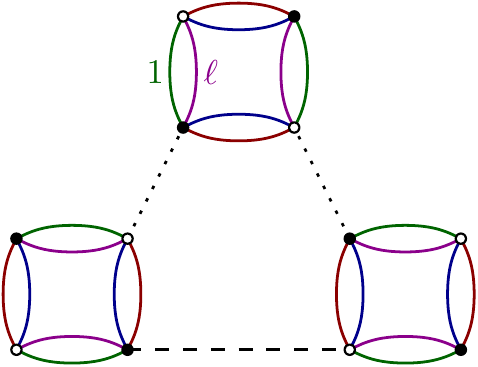}}} \right)\,,
\end{align}

\begin{align}\label{flow_graph_u62}
\partial_k \left( Z(k)^3 u_{6,2} (k) \right) &\sum_{\substack{\ell_1 , \ell_2 =2\\ \ell_1\neq \ell_2}}^4 \vcenter{\hbox{\includegraphics[scale=0.8]{Figures/int62_l.pdf}}} \approx 2 Z(k)^5 k u_4(k) u_{6,2}(k) \sum_{\substack{\ell_1 , \ell_2 =2\\ \ell_1\neq \ell_2}}^4 \left( \vcenter{\hbox{\includegraphics[scale=0.8]{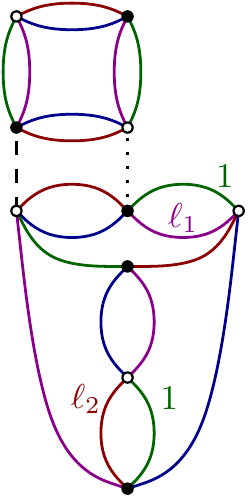}}} + \vcenter{\hbox{\includegraphics[scale=0.8]{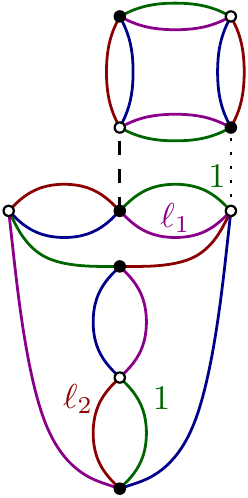}}} \right. \\\nonumber
& \left. + \vcenter{\hbox{\includegraphics[scale=0.8]{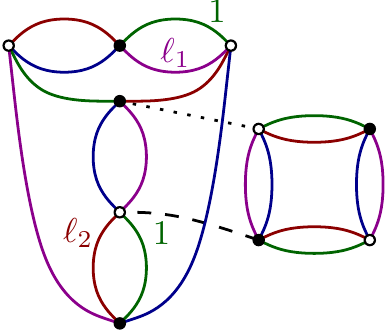}}} + \vcenter{\hbox{\includegraphics[scale=0.8]{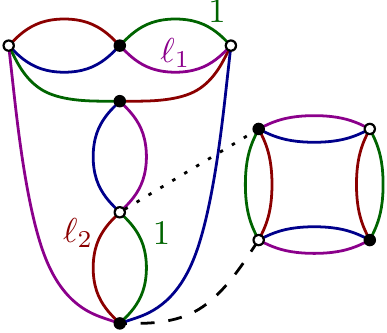}}}  \right)\,.
\end{align}
The derivation of these equations follows a similar route as in \cite{Carrozza:2016tih}; the reader may consult this reference for detailed calculations. These flow equations are approximations, because the right- and left-hand sides do not feature the exact same interaction kernels. To obtain sensible flow equations, we need to "expand" the diagrams on the right-hand sides around the diagrams occurring on the left-hand sides. This intuition is achieved by a Taylor expansion of the corresponding amplitudes. This method relies on the fact that the divergent graphs are tracial. This means that one can evaluate the integrands associated to the one-loop diagrams contributing to the flow around flat bulk holonomies, hence around trivial holonomies (up to gauge). The zeroth orders in such an expansion are then necessarily proportional to bare necklace bubble interactions. The higher orders include derivative couplings and, as a result of our truncation ansatz, will be discarded except for the $2$-point function which must retain second derivatives. 

Let us now illustrate how this works by a simple example. We consider a typical contribution to the right-hand side of equation \eqref{flow_graph_u2}, say $A_l[\phi,\bar{\phi}]$, where $l$ denotes the color paired with color $1$ in the corresponding diagram. It has the following structure: 
\begin{equation}
A_{l}[\phi,\bar{\phi}]=Z^2(k)u_4(k)\int d\textbf{g}_1\textbf{g}_2 dh \bar{\phi}(\textbf{g}_1)\phi(\textbf{g}_2) a_l(g_{11}g_{21}^{-1}h,g_{1l}g_{2l}^{-1}h,h,h)\prod_{i\neq 1,l} \delta(g_{1i}g_{2i}^{-1}) \,,
\end{equation}
where $a_l(g_{11}g_{21}^{-1}h,g_{1l}g_{2l}^{-1}h,h,h)$ is the integrand associated to the diagram loop. We fix $l=2$ for simplicity, and introduce an interpolation between the end group variables: $g_{1k}(t)=g_{1k}\exp(tX_{g_{1k}^{-1}g_{2k}})\,\,, k=1,2$; where $X_g\in \mathfrak{su}(2)$ denotes the smallest Lie algebra element such that $g=\exp (X_g)$. As a result, the Taylor expansion around $t=0$ writes:
\begin{equation}
\int d\textbf{g}_1\textbf{g}_2 \bar{\phi}(\textbf{g}_1)\phi(\textbf{g}_2)\prod_{i\neq 1,2} \delta(g_{1i}g_{2i}^{-1})= \int d\textbf{g}_1\textbf{g}_2\sum_{n_1,n_2} \frac{\bar{\phi}^{(n_1,n_2)}(g_1(t_1=0), g_2(t_2=0),g_3,g_4)}{n_1!n_2!}\phi(\textbf{g}_2)\prod_{i\neq 1,2} \delta(g_{1i}g_{2i}^{-1})\,,
\end{equation}
and graphically:
\beq
\vcenter{\hbox{\includegraphics[scale=0.8]{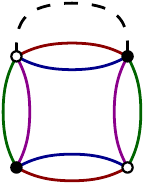}}} \approx \frac{1}{Z(k)} L_1 ( u_2 (k) , u_4 (k) ) \times \vcenter{\hbox{\includegraphics[scale=0.8]{Figures/int2.pdf}}} \; + \frac{1}{6 Z(k) k^2} L_w ( u_2 (k) , u_4 (k) ) \left( \vcenter{\hbox{\includegraphics[scale=0.8]{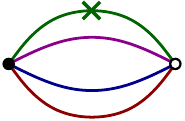}}} + \vcenter{\hbox{\includegraphics[scale=0.8]{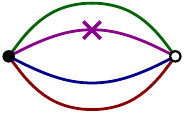}}} \right)\,,
\eeq
where the functions $L_1 ( u_2 (k) , u_4 (k) ) $ and $L_w ( u_2 (k) , u_4 (k) ) $ are defined as (to obtain these formula, we use the translation invariance of the Haar measure over $SU(2)$):
\begin{equation}
\frac{1}{Z(k)}L_1 ( u_2 (k) , u_4 (k) ) :=Z^2(k)ku_4 \int dg_1dg_2 dh\, a_l(g_1,g_2,h,h)\,,
\end{equation}
\begin{equation}
\frac{1}{Z(k)k^2}L_w ( u_2 (k) , u_4 (k) ) := Z^2(k)ku_4 \int dg_1dg_2 dh \, a_l(g_1,g_2h,h,h)\vert X_{g_2}\vert^2\,.
\end{equation}
The same local approximation holds for all the right-hand sides of the flow equations, and typical contractions and relevant loop integrals have been summarized in Table \ref{tableloops}, where the functions $L_w$ and $L_k$ may be exactly computed in the deep ultraviolet sector (the details of the computation may be found in\cite{Carrozza:2016tih}, and we do not reproduce them here). They have the following structure :
\begin{align}\label{L_k}
L_k (u_2 , u_4) &= 2 f_k (u_2 , u_4 ) + \eta (u_2 , u_4 ) g_k (u_2 , u_4) \,.
\end{align}
\begin{align}
L_w (u_2 , u_4) &= 2 f_w (u_2 , u_4 ) + \eta (u_2 , u_4 ) g_w (u_2 , u_4) \,,
\end{align}
where the functions $f_k$, $g_k$, $f_w$ and $g_w$ are given by:
\begin{equation}
f_k (u_2)  = 2 \sqrt{2} \int_{0}^{\infty} dx\frac{x^6\, e^{-x^2} \, \left( 1-e^{-x^2} \right)^{k-1}}{\left( x^2+ u_2(1-e^{-x^2})\right)^{k+1}}\,,
\end{equation}
\begin{equation}
g_k ( u_2)  = 2 \sqrt{2} \int_{0}^{\infty} dx\frac{x^4\, e^{-x^2} \, \left(1-e^{-x^2}\right)^k}{\left( x^2+ u_2(1-e^{-x^2})\right)^{k+1}}\,,
\end{equation}
\begin{equation}
f_w(u_2):=9 \sqrt{2} \int_0^{\infty} \extd x \, e^{-x^2} \left( \frac{x^2}{x^2+ u_2  (1-e^{-x^2})}\right)^2\,,
\end{equation}
\begin{equation}
g_w( u_2) := 9 \sqrt{2} \int_0^{\infty} \extd x \, e^{-x^2}\left(1-e^{-x^2}\right) \left( \frac{x}{x^2+ u_2 (1-e^{-x^2})}\right)^2\,.
\end{equation}
\begin{center}
\begin{tabular}{|R{8.7cm}|C{8.5cm}|}
\hline $ \vcenter{\hbox{\includegraphics[scale=0.6]{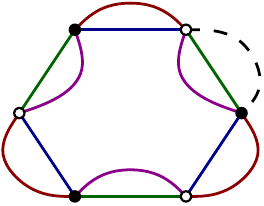}}} \approx \frac{1}{Z(k)} L_1 ( u_2 (k) , u_4 (k) ) \times \vcenter{\hbox{\includegraphics[scale=0.8]{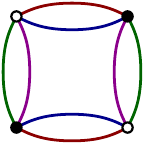}}} $ & $ \vcenter{\hbox{\includegraphics[scale=0.6]{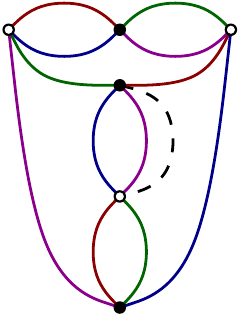}}} \approx \frac{1}{Z(k)} L_1 ( u_2 (k) , u_4 (k) ) \times \vcenter{\hbox{\includegraphics[scale=0.8]{Figures/int4_no-l.pdf}}} $  \\
\hline $ \vcenter{\hbox{\includegraphics[scale=0.6]{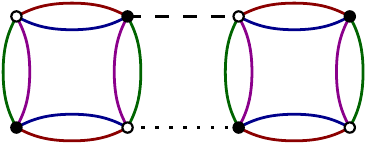}}} \approx \frac{1}{k^2 Z(k)^2} L_2 ( u_2 (k) , u_4 (k) ) \times \vcenter{\hbox{\includegraphics[scale=0.8]{Figures/int4_no-l.pdf}}} $ & $ \vcenter{\hbox{\includegraphics[scale=0.6]{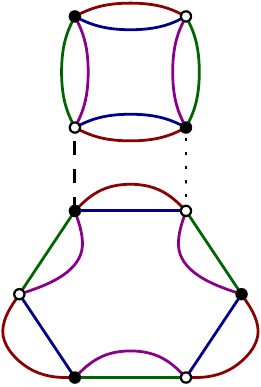}}} \approx \frac{1}{k^2 Z(k)^2} L_2 ( u_2 (k) , u_4 (k) ) \times \vcenter{\hbox{\includegraphics[scale=0.5]{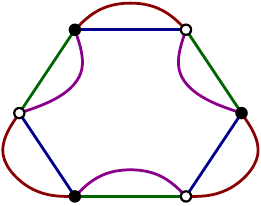}}} $  \\
\hline $ \vcenter{\hbox{\includegraphics[scale=0.5]{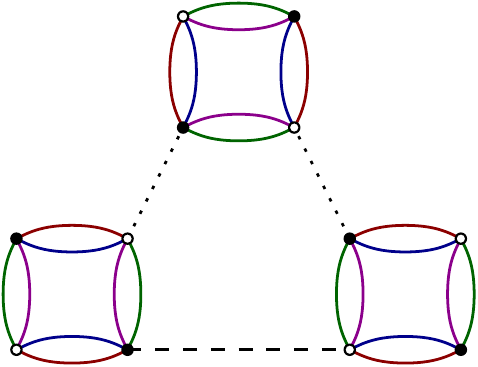}}} \approx \frac{1}{k^4 Z(k)^3} L_3 ( u_2 (k) , u_4 (k) ) \times \vcenter{\hbox{\includegraphics[scale=0.6]{Figures/int61_no-l.pdf}}} $& $ \vcenter{\hbox{\includegraphics[scale=0.5]{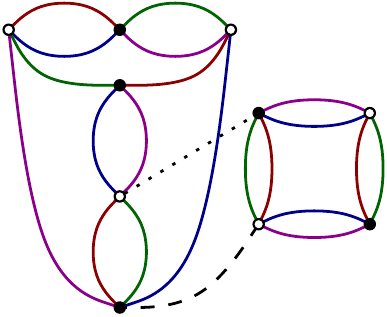}}} \approx \frac{1}{k^2 Z(k)^2} L_2 ( u_2 (k) , u_4 (k) ) \times \vcenter{\hbox{\includegraphics[scale=0.6]{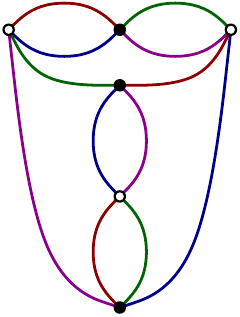}}} $\\
\hline
\end{tabular}
\captionof{table}{Local approximations for all the relevant contractions on the right-hand sides of the flow equations \eqref{flow_graph_u2} to \eqref{flow_graph_u62}.}
\label{tableloops}
\end{center}
Finally, the resulting beta functions in the $\phi^6$ truncation are given by:
\begin{equation}\label{beta_phi6}
\left\lbrace\begin{split}
\beta_{2} &= - \left( 2 + \eta \right) u_2 - 6 L_1 (u_2 , u_4 ) \, u_4 
\\
\beta_4&= - \left( 1 + 2 \eta \right) u_4 - 4 L_1 (u_2 , u_4 ) \, \left(u_{6,1}+4 u_{6,2} \right)+ 4 L_2 (u_2 , u_4 ) \, {u_4}^2 
\\
\beta_{6,1} &= - 3 \eta \, u_{6,1} + 12 L_2 (u_2 , u_4 ) \, u_4 \, u_{6,1} - 6 L_3 (u_2 , u_4 ) \, {u_4}^3 
\\
\beta_{6,2} &= - 3 \eta \, u_{6,2} + 8 L_2 (u_2 , u_4 ) \, u_4 \, u_{6,2} 
\end{split}\right.
\end{equation}
with the anomalous dimension $\eta$
\begin{equation}\label{eta}
\eta =\frac{f_w (u_2) \, u_4}{1 - \frac{1}{2} g_w (u_2) \, u_4}\,.
\end{equation}
The next two sections are devoted to the numerical analysis of this four-dimensional closed autonomous system, focusing in particular on the research of non-Gaussian fixed points. 

\section{Properties of the Gaussian fixed point}\label{sec:gauss}

The Gaussian Fixed Point (GFP) is the origin of the truncated phase space, i.e. the point of coordinates $$\mathrm{GFP}:=(u_2=0,u_4=0,u_{6,1}=0, u_{6,2}=0)\,,$$
at which the beta functions \eqref{beta_phi6} vanish trivially. In this section, we will describe the renormalization group flow in the vicinity of the GFP, around which the perturbative expansion makes sense, and the beta functions are universal, that is, do not depend on the choice of the regularization scheme. Denoting by $\{g_i\}=\{u_2, u_4, u_{6,1}, u_{6,2}\}$ the coupling coordinates, the behavior of the flow in the vicinity of any fixed point $\{g_i^*\}$ is 
\begin{equation}
\beta_i \approx \sum_{j=1}^4 \frac{\partial \beta_i}{\partial  g_j}\bigg\vert_{g_i=g_i^*}\,(g_i-g_i^*)=: \sum_{j=1}^4\beta_{ij}\bigg\vert_{g_i=g_i^*}\,(g_i-g_i^*)\,,
\end{equation}
where $\{\beta_{ij}\}$ is  the \textit{stability matrix}, whose eigenvalues are minus the standard critical exponents. At the GFP, the system \eqref{beta_phi6} leads to:
\begin{equation}[\beta^{\mathrm{GFP}}_{ij}]=\begin{pmatrix}
-2&-6\sqrt{2\pi}&0&0\\
0&-1&-4\sqrt{2\pi}&-16\sqrt{2\pi}\\
0&0&0&0\\
0&0&0&0
\end{pmatrix},
\end{equation}
with eigenvectors $e_1^{\mathrm{GFP}}= (1,0,0,0)^{\mathrm{T}}$, $e_2^{\mathrm{GFP}}=(-6\sqrt{2\pi},1,0,0)^{\mathrm{T}}$, 
$e_3^{\mathrm{GFP}}=(0,0,-4,1)^{\mathrm{T}}$ and $e_4^{\mathrm{GFP}}=(96 \pi,-16 \sqrt{2 \pi},0,1)^{\mathrm{T}}$; they respectively have eigenvalues $-2$, $-1$, $0$ and $0$. As a result, the first two directions are relevant, and the corresponding operators have canonical dimensions $2$ and $1$, whereas the two remaining directions are marginal, with vanishing canonical dimension. The RGE can be numerically integrated, and it is instructive to plot the RG trajectories around the GFP. Expanding the system \eqref{beta_phi6} around the GFP, we find the universal one-loop perturbative flow:
\begin{equation}
  \left\{
      \begin{aligned}
        \beta_2^{(1)} = & -2u_2- 6 \sqrt{2 \pi} u_4 \approx -2 u_2 - 15 u_4\\
        \beta_4^{(1)} = &-u_4 - 4 \sqrt{2 \pi} \left( u_{6,1}+4u_{6,2} \right)- \left( 8 + \sqrt{2} \right) \sqrt{\pi} u_4^2 \approx -u_4 - 10 \left( u_{6,1}+4u_{6,2} \right)- 17 {u_4}^2 \\
        \beta_{6,1}^{(1)} = & - \frac{3}{2} \left( 16 - 7 \sqrt{2} \right) \sqrt{\pi} u_4 u_{6,1}  - 24 \left( \sqrt{11 - 4 \sqrt{6}} - 1\right) \sqrt{2 \pi} {u_4}^3 \approx - 16 u_4 u_{6,1} - 5.8 {u_4}^3\\
        \beta_{6,2}^{(1)} =&-\left( 16- \frac{5}{2} \sqrt{2} \right) \sqrt{\pi} u_4 u_{6,2} \approx - 22 u_4 u_{6,2}\\
      \end{aligned}
    \right.
\end{equation}
Note that, in addition to one-loop linear terms coming from the marginal couplings for $\beta_4$ and $\beta_{6,2}$, we have retained the quartic and cubic one-loop contributions involving $u_4$. Figure \ref{VGFP} represents these trajectories in the planes $u_2=u_{6,2}=0$ and $u_2=u_{6,1}=0$. Interestingly, in both cases, the trajectories in the upper part of the diagrams, for positive marginal couplings in the infrared are necessarily repelled from the GFP in the ultraviolet, and then, the respective couplings do not go to zero in this limit, meaning that the model in not asymptotically free (note that in the diagrams of Figure \ref{VGFP}, the arrows are oriented toward the infrared). This conclusion, and the qualitative structure of the vector field are very reminiscent of the non-Abelian three-dimensional model considered in\cite{Carrozza:2016tih}. 

\begin{center}
\includegraphics[scale=.5]{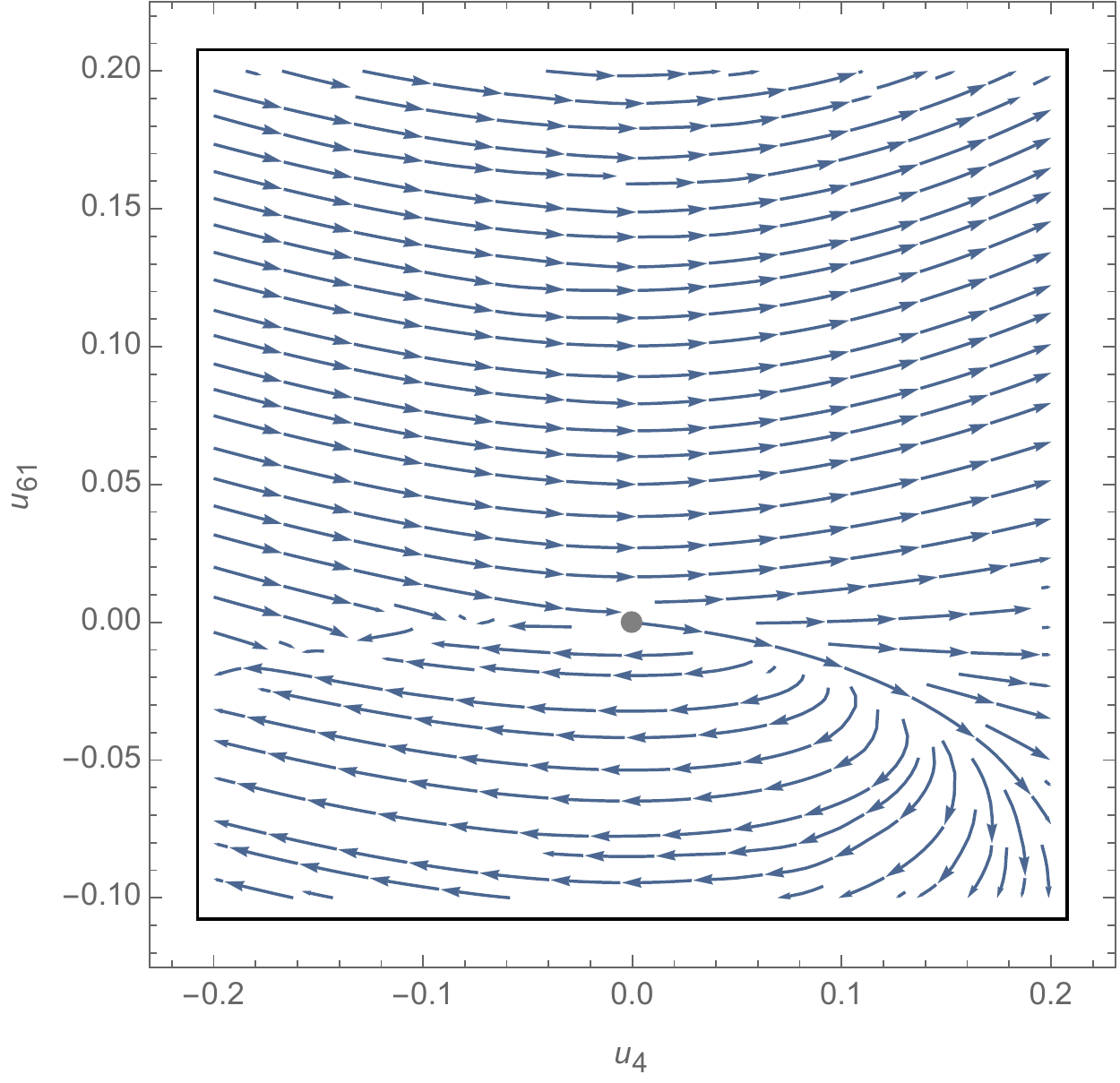}
\includegraphics[scale=.5]{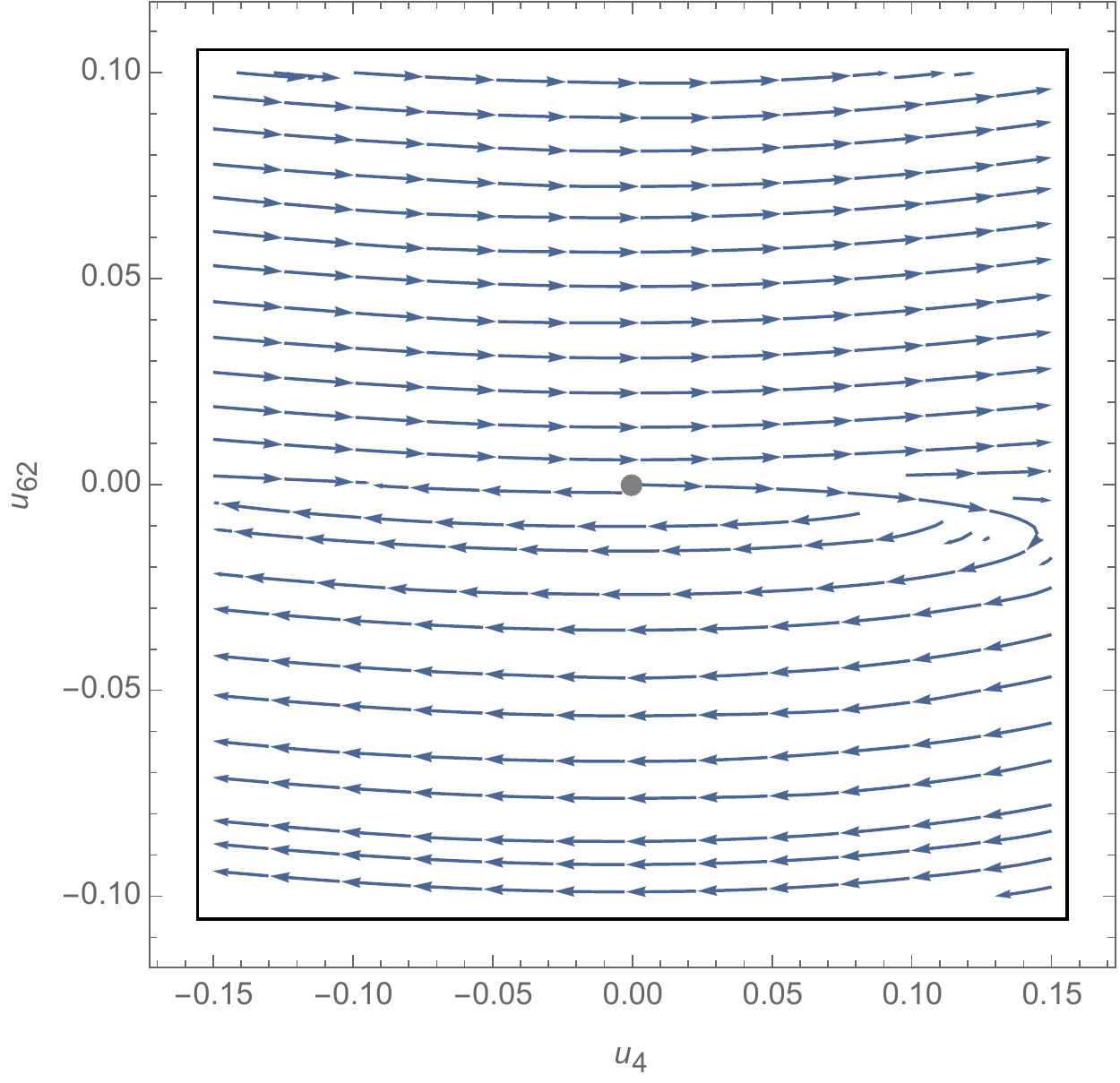}
\captionof{figure}{Phase diagrams around the Gaussian fixed point in the planes $u_2=u_{6,2}=0$ (on the left) and $u_2=u_{6,1}=0$ (on the right).}\label{VGFP}
\end{center}

\section{Non--perturbative regime}\label{sec:ng-fp}

\subsection{Line of fixed points}
Jut like for the three-dimensional TGFT considered in \cite{Carrozza:2016tih}, the system \eqref{beta_phi6} admits a one-parameter family of non-Gaussian fixed points. Indeed, it is easy to see that the line :
\begin{equation}
u^*(s):=(u_2^*,u_4^*,u_{6,1}^*,u_{6,2}^*)=(0,0,s,-s/4)\,,
\end{equation}
is a fixed point for any $s\in \mathbb{R}$, the special point $u^*(s=0)$ corresponding to the GFP. The authors in \cite{Carrozza:2016tih} have shown that this line of fixed points disappears in higher truncations, and therefore seems to be a pathological feature of the crude truncation to $\phi^6$ interactions, with no particular physical relevance. We assume that the same conclusion holds for the model of the present, and discard this line of fixed points of our analysis, focusing on \textit{isolated fixed points}.

\subsection{Isolated fixed point}\label{IFP6}

In addition to this pathological line of fixed points, Mathematica allows to compute solutions of the fixed point equations, providing isolated fixed points. The technical difficulties are the same already encountered in \cite{Carrozza:2016tih} for the three-dimensional non Abelian melonic model. Due to the complicated $u_2$-dependence of the loop integrals $L_n(u_2,u_4)$, a direct use of the built-in Mathematica numerical equation solvers do not work, and we have to rely on piecewise polynomial interpolations in order to select candidate fixed point values of $u_2$. In this way, we find one isolated fixed point, associated to a positive value of $u_2$. 
\begin{figure}[h]
\centering
\includegraphics[scale=.5]{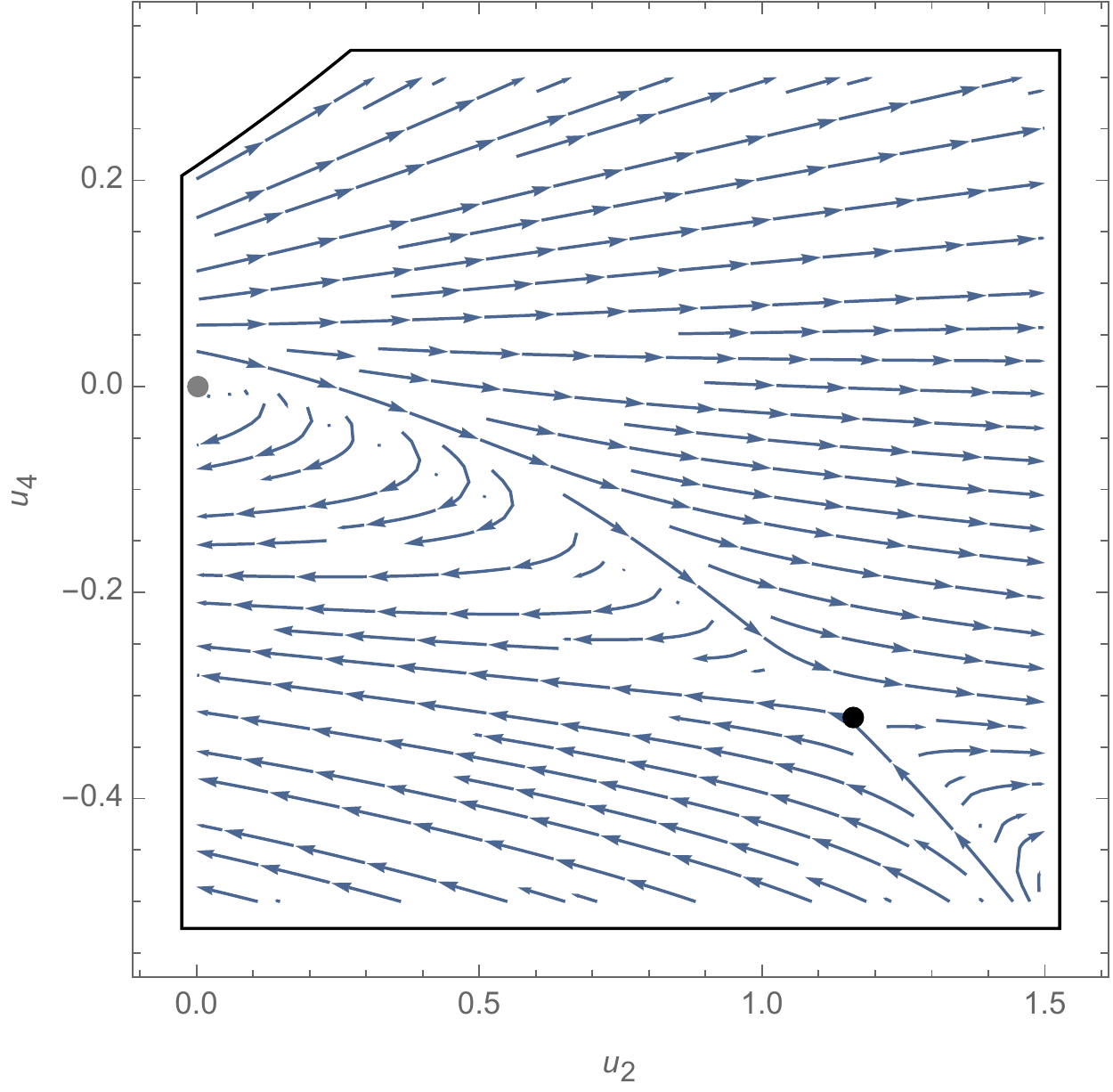}
\includegraphics[scale=.5]{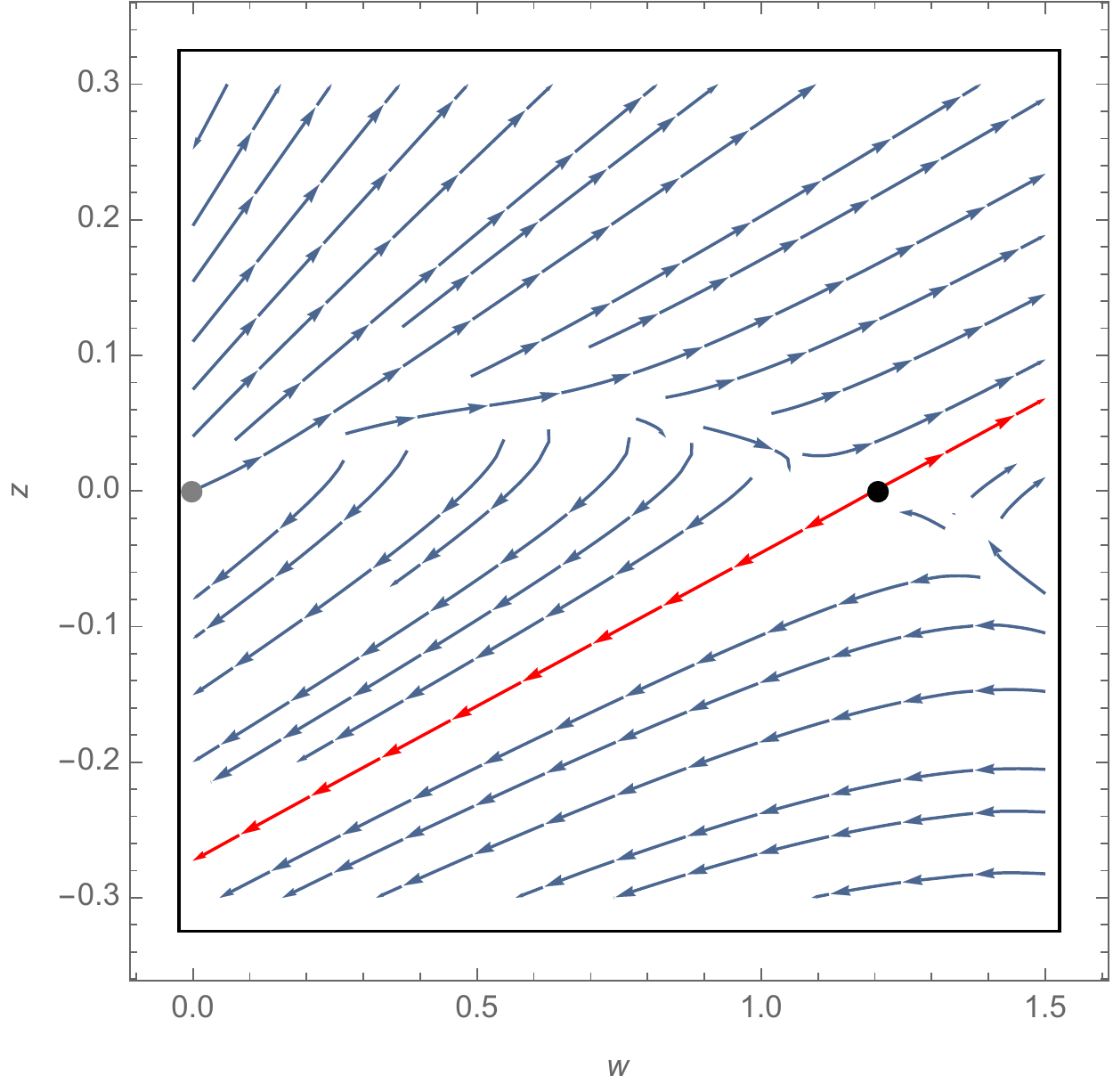}
\captionof{figure}{Two-dimensional projections of the four-dimensional truncated renormalization group flow. In both cases, the black (resp. grey) dots represents the non-Gaussian fixed point IFP (resp. the Gaussian fixed point), and as for the previous diagrams, the arrows point toward the infrared. On the left: projection onto the plane $\{u_{6,2}=0, u_{6,1}=u_{6,1}^*\}$; the boundary in the upper left corner being the singularity $X=0$. On the right: projection onto the plane containing the origin, the non-Gaussian fixed point, and its relevant direction $\textbf{V}$; $w$ is a coordinate along the line connecting the two fixed points; $z$ is a parameter along the orthogonal direction; the red trajectories arise from $\mathrm{IFP}$ in the directions $\pm \textbf{V} $.}\label{Twodimproj}
\end{figure}
\noindent
Note that the denominator of $\eta$ in equation \eqref{eta}:
\beq
X(u_2,u_4):=1-1/2g_w(u_2)u_4\,,
\eeq 
introduces a singularity surface where our parametrization of the flow breaks down. Since the Gaussian fixed point $\mathrm{GFP}$ is on the side $X>0$, one cannot trust computations in the region $X<0$. 

The numerical analysis shows the existence of a unique isolated fixed point in the region $X>0$, for the value:
\begin{equation}
\mathrm{IFP} = \left( u_2^*, u_4^*, u_{6,1}^*, u_{6,2}^* \right) \approx \left( 1.2 , - 0.32 , -0.013 , 0 \right)\,.
\end{equation}
The anomalous dimension at $\mathrm{IFP}$ is $\eta \approx -0.71$. The characteristics of this fixed point seem to indicate that it plays the role of an ultraviolet fixed point. Indeed, computing the eigenvalues of the stability matrix, we find the diagonalization $\beta_{\mathrm{diag}}\approx ( - 2.2 , 0.33 , 1.53 , 1.53)$, meaning that $\mathrm{IFP}$ has one (infrared) relevant direction and three irrelevant directions. \\

\noindent
Two slices of the RGE trajectories are represented on Figure \ref{Twodimproj}, and highlight the influence of the fixed point IFP. The rightmost Figure corresponds to a projection of the RG flow into the two-dimensional plane containing both the Gaussian fixed point, IFP, and the eigenvector  $\textbf{V}\approx(0.1, -0.05, 0, 0)$ with critical exponent $\approx 2.2$; and suggests the existence of two distinct low energy phases separated by the integral surface spanned by irrelevant directions around the fixed point IFP.

\section{Higher order truncations}

As explained in the previous section \ref{sec:frg}, one of the main challenges for the FRG program is the reliability of the predictions in a given approximation scheme, that is, in a given truncation and for a given regulator. In standard field theory over manifolds, the first step towards a definitive conclusion about the existence of a non-trivial fixed point is to test the robustness in higher order truncations, at least in a restricted sector of the theory space. The convergence and stability of the critical exponents are expected to be a good criterion for the quality of the truncation (see \cite{Reuter:2012id} for a discussion in the context of Einstein quantum gravity and \cite{Carrozza:2016tih} in the TGFT context). In this Section, we investigate the RGE equations \eqref{WM} in a restricted sector of the theory space around the fixed point $\mathrm{IFP}$, providing a first indication on the robustness of our predictions, as well as on their limitations.

\subsection{Non-branching theory space}

Because the number of necklace graphs grows exponentially with the number of nodes, a systematic exploration of the theory space in higher order truncations remains a difficult challenge. However, in the $\phi^6$ truncation, we have showed that the necklace of type $(6,2)$ does not contribute to $\mathrm{IFP}$, suggesting to look at a restricted family of necklace graphs for higher order investigations. Note that a similar mechanism was found by the authors of \cite{Carrozza:2016tih} in the melonic sector, and this section is largely inspired from this earlier investigation. The authors have called \textit{non-branching} the restricted family of melons found in rank three. In the same way, we will call \textit{non-branching necklace} the stable restricted sector of the whole theory space on which we will focus. \\
\begin{figure}
\centering
\includegraphics[scale=0.8]{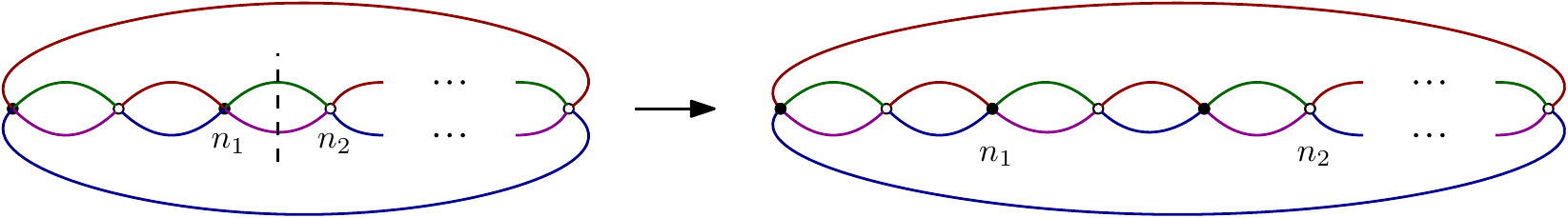} 
\caption{$2$-dipole insertion between two nodes $n_1$ and $n_2$.}\label{genpro}
\end{figure}
\begin{figure}
\centering
\includegraphics[scale=.8]{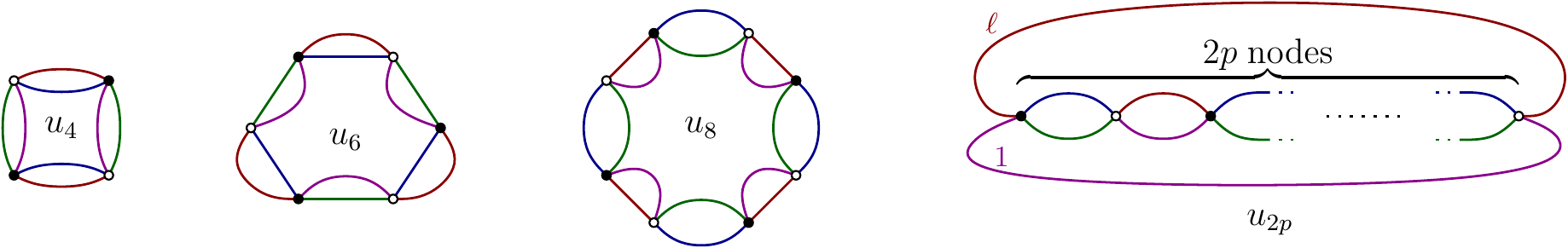}
\caption{Non-branching necklace bubbles and their associated coupling constants.}\label{non-branching}
\end{figure}

Similarly to the non-branching melons, non-branching necklaces are defined through recursive insertions of $2$-dipoles in place of two lines connecting two black and white nodes. Figure \ref{genpro} describes the general procedure while Figure \ref{non-branching} shows the first non-branching bubbles as well as the general structure for $2p$ nodes. \\

\noindent
As anticipated above, the most exciting property of the non-branching necklaces comes from their stability under the renormalization group in the deep ultraviolet limit. This comes from the fact that the local approximation of a necklace graph involving only non-branching necklaces remains a non-branching necklace. Then, we consider the following infinite-dimensional ansatz for the effective average action on the non-branching necklace bubbles subspace:
\begin{align}\label{ansatz_non-branching}
\Gamma_k &= -Z(k) \sum_{\ell=1}^4 \vcenter{\hbox{\includegraphics[scale=0.8]{Figures/int2_laplace.pdf}}}  + Z(k) k^2 u_2(k) \vcenter{\hbox{\includegraphics[scale=0.8]{Figures/int2.pdf}}} \\
& \qquad + \sum_{p \geq 2} Z(k)^p k^{3-p} \frac{u_{2p} (k)}{p}  \sum_{\ell=2}^4 \vcenter{\hbox{\includegraphics[scale=0.8]{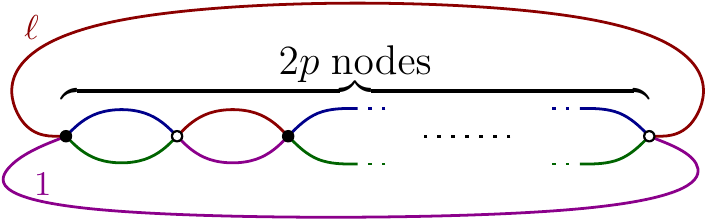}}}  \nn\,.
\end{align}   
Note that the factor $1/p$ takes into account the number of automorphisms of non-branching necklaces with $2p$ nodes. The derivation of the flow equations follows the same strategy as for the $\phi^6$ truncation. A graph with a loop of length $k$ involves a loop integral weight function $L_k(u_2,u_4)$, as a specific combinatorial factor taking into account the number of contractions providing the same interaction bubble. Up to a factor $2$, the computation of this factor follows the melonic computation, whose details may be found in \cite{Carrozza:2016tih}. Note that the factor $2$ comes from the fact that we have twice as many contractions per graphs than in the melonic sector. Then, the beta function $\beta_{2p}$  for any $u_{2p}$ writes as:
\beq
\beta_{2p} = \left( p - 3 - p \eta \right) u_{2p} + 2 p \sum_{k = 1}^p (-1)^{k} L_{k} (u_2, u_4) \sum_{ \{ n_{2q} \} \in \mathcal{D}_{k,p} 
}
{k\choose{\{ n_{2q} \}} }  \prod_{q \geq 2} (u_{2q})^{n_{2q}}
\,, \qquad \forall p \geq 2 \,,
\eeq
where
\beq
\mathcal{D}_{k,p} := \Big\{ \{ n_{2q} \vert q \in \mathbb{N}  , q \geq 2\} \Big\vert \underset{q \geq 2}{\sum} n_{2 q} = k \; \mathrm{and} \; \underset{q \geq 2}{\sum} q n_{2 q} = p+k\Big\}
\eeq
and
\beq
{k\choose{\{ n_{2q} \}}} = \frac{k! }{ \underset{q\geq2}{\prod} (n_{2q}!)}
\eeq
is the standard multinomial coefficient.
Evaluating explicitly the first beta functions in this infinite tower of flow equations, we find:
\begin{equation}
\left\lbrace\begin{split}
\beta_{2} &= - \left( 2 + \eta \right) u_2 -6 L_1 (u_2, u_4) \, u_4 
\\
\beta_4&= - \left( 1 + 2 \eta \right) u_4 - 4 L_1 (u_2, u_4) \, u_{6} +4 L_2 (u_2, u_4) \, {u_4}^2 
\\
\beta_{6} &= - 3 \eta \, u_{6} - 6 L_1 (u_2, u_4) \, u_8 + 12 L_2 (u_2, u_4) \, u_4 u_6 - 6 L_3 (u_2, u_4)  {u_4}^3 \\
\beta_{8} &= \left( 1 - 4 \eta \right) \, u_{8} - 8 L_1 (u_2, u_4) \, u_{10} + 8 L_2 (u_2, u_4) \, \left( 2 u_4 u_8 + {u_6}^2 \right) \\ 
& \qquad - 24 L_3 (u_2, u_4) \, {u_4}^2 u_6 + 8 L_4 (u_2, u_4)  \, {u_4}^4 
\\
\beta_{10} &= \left( 2 - 5 \eta \right) \, u_{10} - 10 L_1 (u_2, u_4) \, u_{12} + 20 L_2 (u_2, u_4) \, \left( u_4 u_{10} + u_6 u_8 \right) \\ 
& \qquad - 30 L_3 (u_2, u_4)  \, \left( {u_4}^2 u_8 + u_4 {u_6}^2 \right)+ 40 L_4 (u_2, u_4)  \, {u_4}^3 u_6 - 10 L_5 (u_2, u_4) \, {u_4}^5 \\
\beta_{12} &= \left( 3 - 6 \eta \right) \, u_{12} - 12 L_1 (u_2, u_4) \, u_{14} + 12 L_2 (u_2, u_4) \, \left( 2 u_4 u_{12} + 2 u_6 u_{10} + {u_8}^2 \right) \\ 
& \qquad - 12 L_3 (u_2, u_4)  \, \left( 3 {u_4}^2 u_{10} + 6 u_4 u_6 u_8 + {u_6}^3 \right)+ 24 L_4 (u_2, u_4)  \, \left( 2 {u_4}^3 u_8 + 3 {u_4}^2 {u_6}^2 \right) \\
& \qquad - 60 L_5 (u_2, u_4) \, {u_4}^4 u_6 + 12 L_6 (u_2, u_4) \, {u_4}^6 \\
& \cdots
\end{split}\right.
\end{equation}

\subsection{Isolated fixed point}
As announced, the restriction to the non-branching family allows investigations in high order truncations. The results for truncations up to order $12$ in a local potential approximation have been summarized in Table \ref{IFP_non-branching}. They confirm the existence of an ultraviolet fixed point with positive effective mass $u_2$. Indeed, the  fixed point $\mathrm{IFP}$ is reproduced in all truncations (up to order $12$), with a single infrared relevant direction whose critical exponent varies from $2.1$ to $2.5$, and seems to confirm the picture of Section \ref{IFP6} about the existence of two infrared phases separated by the surface locally generated by irrelevant directions. \\

\begin{table}[ht]
\centering
\begin{tabular}{|c||c|c|c|c|c|}
\hline
 Truncation order & $4$ & $6$ & $8$ & $10$ & $12$ \\ \hline\hline
$u_2^*$ & $0.93$ & $1.2$ & $1.4$ & $1.7$ & $2.3$ \\ \hline
$u_4^*$ & $-0.23$ & $-0.32$ & $-0.42$ & $-0.58$ & $-1.0$ \\ \hline
$u_{6}^*$ & -- & $-0.013$ & $-0.035$ & $-0.088$ & $-0.37$ \\ \hline
$u_{8}^*$ & -- & -- & $-0.0053$ & $-0.026$ & $-0.23$ \\ \hline
$u_{10}^*$ & -- & -- & -- & $-0.0072$ & $-0.15$ \\ \hline
$u_{12}^*$ & -- & -- & -- & -- & $-0.082$ \\ \hline
$\theta_1$ &  $2.1$ & $2.2$ & $2.3$ & $2.4$ & $2.5$ \\ \hline
$\theta_2$ & $-0.42$ & $-0.33$ & $-0.25$ & $-0.16$ & $-0.11$ \\ \hline 
$\theta_3$ & -- & $-1.5$ & $-1.3$ & $-1.3$ & $-1.3$ \\ \hline
$\theta_4$ & -- & -- & $-3.4$ & $-3.2$ & $-3.4$ \\ \hline
$\theta_5$ & -- & -- & -- & $-5.5$ & $-5.6$ \\ \hline
$\theta_6$ & -- & -- & -- & -- & $-8.0$ \\ \hline
$\eta$ & $-0.64$ & $-0.71$ & $-0.77$ & $-0.84$ & $-0.95$ \\ \hline
\end{tabular}
\caption{Properties of $\mathrm{IFP}$ in the non-branching necklace truncation, up to order $12$.}\label{IFP_non-branching}
\end{table}

\noindent
Note that, as pointed out in the melonic sector in \cite{Carrozza:2016tih}, all the critical exponents remain stable over the different truncation, but do not converge, even if the amplitude of the variations seems to be smaller for necklaces. As for the melons, we conjecture that this is a pathology of the local potential approximation, which discards derivative couplings occurring in the local expansion of the loop contractions on the right-hand side of the flow equations (see \cite{Morris:1994ie} for a general discussion of the derivative expansion). We expect a much better convergence of the critical exponents under refinement of the truncation, if these derivative couplings are systematically taken into account.

\section{Conclusions}\label{sec:conclu} 
We have introduced a rank-four TGFT model based on the group manifold $\SU(2)$ and characterized by necklace interactions up to order 6. We have proven that the model is renormalizable at all orders in perturbation theory. This is possible because the canonical dimension associated to the different interactions of the model is such that the renormalization flow of melonic interactions, which would be non-renormalizable in this case, leads to their suppression (compared to necklace interactions) in the ultraviolet regime. We have investigated the detailed properties of the Gaussian fixed point using functional renormalization group techniques, which we then also used to identify non-perturbatively other non-trivial UV fixed points of the model. The analysis has also been extended to higher truncations, in the special case of non-branching interactions. 

\noindent These results show that TGFTs, in contrast to the simpler tensor model, can naturally go beyond the melonic sector of interactions simply thanks to their renormalization group flow, in turn originated by the richer combinatorial structure of their interactions, without any fine-tuning. This may suggest also a richer phase diagram for the same models, which however requires much more work to be elucidated in detail. Importantly, the $\SU(2)$ model we have analyzed is expected to have a structure of divergences quite similar to that of 4d quantum gravity models. Therefore, our results constitute one more step towards a renormalizability analysis of these physically more interesting GFT models.

\addtocontents{toc}{\protect\setcounter{tocdepth}{1}}
\addcontentsline{toc}{section}{References}
\bibliographystyle{JHEP}
\bibliography{biblio}

\end{document}